\documentclass[aps,twocolumn,showpacs,superscriptaddress,preprintnumbers,floatfix,amsmath,amssymb,eufrak,table,nofootinbib]{revtex4}
\usepackage{ulem}
\usepackage{color}
\usepackage{graphicx}  
\usepackage{appendix}
 \usepackage{epsfig}
\usepackage[colorlinks=true,linkcolor= red, citecolor = cyan, urlcolor = blue ]{hyperref}
\usepackage{epstopdf}
\usepackage{amsmath}
\usepackage{subfig}
\usepackage{comment}
\usepackage{float}
\usepackage{multirow}
\usepackage{ragged2e} 

\newcommand{\be}{\begin{equation}}
	\newcommand{\ee}{\end{equation}}
\newcommand{\ba}{\begin{eqnarray}}
	\newcommand{\ea}{\end{eqnarray}}

\begin{document}
\title{Charmonium suppression in fixed target proton-nucleus collisions}
	
\author{Sourav Kanti Giri}\email{sk.giri@vecc.gov.in}
\affiliation{Variable Energy Cyclotron Centre, 1/AF Bidhan Nagar, Kolkata 700 064, India}
\affiliation{Homi Bhabha National Institute, Mumbai - 400085, India}
\author{Partha Pratim Bhaduri}\email{partha.bhaduri@vecc.gov.in}
\affiliation{Variable Energy Cyclotron Centre, 1/AF Bidhan Nagar, Kolkata 700 064, India}
\affiliation{Homi Bhabha National Institute, Mumbai - 400085, India}\email{partha.bhaduri@vecc.gov.in}
\author{Biswarup Paul}\email{biswarup.babu@gmail.com}
\affiliation{Bali Ram Bhagat College, Samastipur - 848101, Bihar, India}
\author{Santosh K. Das}\email{santosh@iitgoa.ac.in}
\affiliation{School of Physical Sciences, Indian Institute of Technology Goa, Ponda-403401, Goa, India}%
\date{\today}

\begin{abstract}
In this article, we perform a systematic investigation of the cold nuclear matter (CNM) effects, operative on charmonium ($J/\psi$, $\psi(2S)$) production, in fixed target proton-nucleus (p+A) collisions. Influence on charmonium production cross section due to the interplay of three different plausible CNM effects namely the initial-state parton energy loss, nuclear shadowing, and final-state absorption of the resonant states, are evaluated in detail. The available data on charmonium production in fixed target p+A collision experiments from SPS, Fermilab and HERA-B are examined for this purpose. The beam energy dependence of the observed $J/\psi$ production patterns are utilized to anticipate level of "normal" absorption in the upcoming proton induced collisions by the NA60+ experiment at CERN SPS and the CBM experiment at FAIR SIS100 accelerator facilities.
\end{abstract}
	
\maketitle

\section{Introduction}

The bound states of charm ($c$) and anti-charm ($\bar{c}$) quarks stable under strong decay are collectively called charmonia. In relativistic nucleus-nucleus (A+A) collisions, suppressed production of charmonium states has been investigated for a long time, as a diagnostic and experimentally viable signature to indicate the formation of quark-gluon plasma (QGP) in the laboratory~\cite{Matsui:1986dk,Vogt:1999cu,Kluberg:2009wc,Scomparin:2016gog,Bhaduri:2020yea}. However, to identify the genuine "anomalous" suppression pattern induced by a hot and dense medium in ion-ion collisions and connect it to the onset of color deconfinement, a precise understanding of the production dynamics of the charmonium states in proton-nucleus (p+A) collisions is an essential prerequisite. In these collisions, the production gets affected inside the target nucleus, due to the presence of several effects of different physics origin and coming into play at the different stages of resonance formation, dubbed collectively  as cold nuclear matter (CNM) effects. Consequently, the measured resonance production cross sections increase less than linearly with the mass number (A) of the target, as observed in the data. In fixed target proton-induced collisions, measurement of $J/\psi$ suppression was carried out by the NA3~\cite{NA3:1983ltt}, NA38~\cite{NA38:1998lyg,NA38:1998udo}, NA51~\cite{NA51:1998uun}, NA50~\cite{NA50:2003pvd, NA50:2006rdp, NA50:2004sgj} and NA60~\cite{NA60:2010wey} experiments at the SPS, E772~\cite{Alde:1990wa}, E866~\cite{NuSea:1999mrl} and E906~\cite{R.Shahoyan} experiments at Fermilab, HERA-B~\cite{HERA-B:2008ymp} experiments at DESY, over a wide range of beam energy starting from 120 GeV to 920 GeV and for a variety of nuclear targets. In the so-called $\alpha$-parametrization, the nuclear effects are quantified by comparing the yields of different nuclear targets in a certain kinematic domain probed by the particular measurement and fit the target mass dependence following a power law: $\sigma_{pA} = \sigma_{0} A^{\alpha}$, where $\alpha < 1 $ indicates suppression. A more common practice is to analyze the data in the Glauber model framework~\cite{Miller:2007ri} and express the target mass dependence through an effective "absorption" cross section, $\sigma_{eff}^{J/\psi}$, quantifying the overall nuclear dissociation effects operative on the resonance states throughout their evolution. $\sigma_{eff}^{J/\psi}$, as extracted from the data measured in p+A collisions at the SPS by the NA50 and subsequently the NA60 experiments exhibited a clear dependence on collision energy, with stronger suppression at lower beam energy. It is well known by now that in p+A interactions, charmonium production is influenced by a multitude of processes induced by the cold nuclear matter. Depending on their time of occurrence during the course of the resonance formation, CNM effects can be broadly classified into two categories. In literature most of the models theorize charmonium production in hadronic collisions as a factorizable two step process. Production of an initially compact color octet $c\bar{c}$ pair in hard ($Q^{2} \ge 4m_{c}^{2}$, where $m_{c} = 1.3$ GeV/$c^2$ is the charm quark mass) collisions is the first stage and can be described by the perturbative QCD (pQCD). The second stage is the formation of the color neutral bound states. The process being soft, is non-perturbative in nature and cannot be described by the first principle QCD calculations. Different models have been formulated in literature to account for the hadronization of the evolving $c\bar{c}$ pairs to color singlet physical resonances. In p+A (also in A+A) collisions initial state effects occur before the birth of a heavy quark pair and thus modify the $c\bar{c}$ perturbative production cross section. Modification of parton flux inside the target nucleus, as encoded in nuclear parton densities or the energy loss of the projectile partons due to multiple scattering inside the target, can modify the $c\bar{c}$ production cross section as compared to elementary p+p collisions. On the other hand, the final state CNM effects become operative on the nascent $c\bar{c}$ pairs in the pre-resonant or resonant stage. Different final state effects so far studied in literature include nuclear absorption~\cite{Kharzeev:1996yx}, multiple scattering~\cite{Qiu:1998rz,Bhaduri:2011ht} or energy loss~\cite{Arleo:2012hn} of the $c\bar{c}$ pairs via gluon radiation while traversing out of the nuclear matter of the target. The relative influence of different effects essentially depends on the collision energy, kinematic window probed by a particular measurement, together with the size of the target nucleus. Medium induced soft gluon radiation, causing energy loss of the expanding color octet $c\bar{c}$ pairs is the main effect effect in the final stage when the pair retains its color on its entire path inside the nuclear medium and $J/\psi$ hadronization occurs outside the target nucleus. Alternatively, if the color neutralization and/or hadronization occurs inside the target, nuclear absorption is more suitable scenario to describe  the charmonium suppression in p+A collisions. Usually at lower collision energies, the $J/\psi$ mesons are produced with low momentum. Hence their formation length evaluated in the rest frame of the target is smaller than the distance covered by the $c\bar{c}$ pairs from their point of production till they exit target nucleus. This causes  nuclear absorption as the dominant source of final state dissociation. High momentum $J/\psi$ mesons would traverse nuclear medium as colored $c\bar{c}$ pairs and color neutralization occurs outside the target. $J/\psi$ data collected at very high collision energies and/or high forward rapidities are thus generally not suitable to fit within the final state absorption scenario. In Ref.~\cite{Arleo:2006qk} the authors analyzed all the then available hadroproduction and leptoproduction mid-rapidity $J/\psi$ data on nuclear targets within the Glauber model framework. A global fit to all data, irrespective of the measurements collected with different beam particles and energies, indicated a single universal $J/\psi-N$ inelastic cross section, $\sigma_{J/\psi-N} = 3.4 \pm 0.2$ mb. Similar analysis including nuclear effects on parton densities yielded similar results. In absence of any strong nuclear modifications of the parton densities, no significant energy dependence of the $J/\psi–N$ interaction was observed. A reanalysis of nuclear absorption cross section with the PHENIX d+Au measurements using the same model framework with  the EPS08 nuclear parton densities was presented in  Ref.~\cite{Tram:2008zz}. The strongest suppression as was reported in the PHENIX analysis, led to an enhanced absorption cross section from $3.5 \pm 0.3$ mb to $5.4 \pm 2.5$ mb, derived with the free proton PDF. Stronger modification of gluon densities in the EPS08 model as compared to the EKS98 nPDF scheme was found to result in larger absorption cross section at fixed target energies while a smaller nuclear absorption at RHIC. An improved analysis of the cross sections of $J/\psi$ mesons produced in p+A collisions in fixed target experiments in the energy domain from 200 to 920 GeV and in $\sqrt{s_{\rm NN}} = 200 $ GeV d$+$Au collisions at RHIC, was carried out in Ref.~\cite{Lourenco:2008sk}. Employing different models of parton distributions with and without nuclear modifications and Glauber formalism to model the final state dissociation, the study revealed a significant dependence of the dissociation cross section on collision energy and the kinematics of the $J/\psi$ mesons under consideration. The level of final state dissociation of mid-rapidity $J/\psi$ mesons, significantly decreases with beam energy, a feature qualitatively observed for all the investigated parton distribution models. However, the numerical values of the extracted $\sigma_{J/\psi-N}$ were found to depend on the specific PDF set. A strong enhancement of gluon densities in the initial state required a larger value of the final state absorption cross section, as compared to a free proton PDF, to describe the measured suppression pattern present in the data. The role of the parton shadowing in $J/\psi$ production in p+A and A+A collisions at SPS was further investigated in Ref.~\cite{Arnaldi:2009it, Bhaduri:2014daa}. \\

In addition to initial state shadowing or final state absorption, charmonium production in p+A collisions might also be affected due to the initial state energy loss of the projectile partons inside the target nucleus. In Ref.\cite{Vogt:1999dw}, the $x_{F}$ dependence of charmonium production was studied in hadron-nucleus collisions. Various nuclear effects such as shadowing of the target parton distributions, energy loss of the beam partons, final-state absorption, interactions with the co-movers and intrinsic heavy-quark components were studied separately and incorporated into a model of charmonium suppression. The model results were compared to the then available preliminary data from E866 experiment. Apart from that, the impact of the initial state energy loss on the charmonium production in p+A collisions is not very widely studied at fixed target energies. Due to soft multiple scatterings inside the target nucleus, the incoming quarks and gluons of the beam proton may loose energy, before they undergo hard scattering with the target partons to produce a $c\bar{c}$ pair. As a result the overall $c\bar{c}$ production cross section may get modified. The goal of our present article is to study the influence of the beam parton energy loss on the final state absorption of the charmonium states. Suitable data corpus available for the inclusive $J/\psi$ and $\psi(2S)$ production cross sections, in the fixed target p+A collisions are analyzed for this purpose. Unlike previous studies, we do not analyze all the available data on charmonium production from different fixed target experiments. Rather we calculate the average formation length of the $J/\psi$ mesons in the target rest frame within the kinematic domain specific to each measurement and opt the data sets which fit well within the nuclear absorption scenario. \\

The remaining of the article is organized as follows: in section II, we provide a brief formalism for the calculation of differential charmonium production cross section, along with the contributions from the various nuclear effects in p+A collisions. Section III offers a summary and justification for the selection of the analyzed experimental data. The results obtained are discussed in section IV. Finally, section V presents the summary and conclusions of our present study.

\section{The formalism for J/$\psi$ production in cold nuclear medium}
\begin{figure}
    \includegraphics[width=1.0\linewidth]{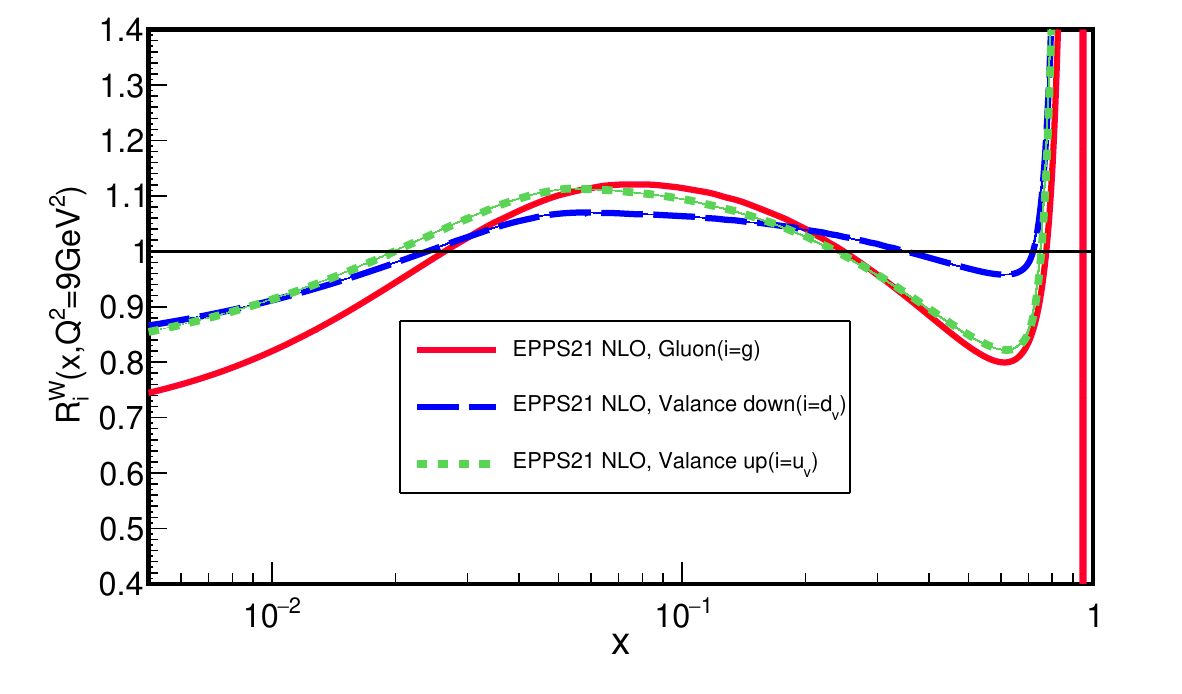}
    \caption{$x$ dependence of the shadowing parameter for gluons, valance up quarks and valence down quarks inside W nucleus, evaluated using EPPS21NLO nPDF package at a momentum scale $Q=3.1$ GeV, suitable for $J/\psi$ production.}
    \label{Fig: pdf}
\end{figure}
 In the literature different models have been proposed to describe the  charmonium production in hadronic collisions~\cite{Schuler:1994hy,Schuler:1996ku,Lansberg:2002cz,Bodwin:2012ft,Chen:2021tmf}. For the present study, we adopt the Color Evaporation Model (CEM)~\cite{Gavai:1994gb,Gavai:1994in} augmented with the initial and final state CNM effects, to calculate $J/\psi$ production in p+A collisions. In hadro-production, CEM treats all quarkonium states identically to the open heavy flavor hadron production, with the condition that the upper limit of the invariant mass of the heavy quark pair is constrained to be less than twice the mass of the lightest meson formed by one of the heavy quark constituents. According to this model, the production cross section of the $i^{th}$ charmonium state is an energy and kinematics independent constant fraction, denoted as $F_{i}$, of the total closed $c\bar c$ production. The invariant mass ($m$) of the closed $c\bar c$ pairs are restricted to lie within the range from $2m_{c}$ to $2m_D$, where $m_c=1.3$ GeV and $m_D =1.87$ GeV are the masses of the charm quark and the lightest D meson, respectively. For p+A collisions, the incorporated initial state effects include the nuclear modifications through the nuclear parton distribution functions (nPDF) inside the target and the energy loss of the incoming beam partons based on the standard parameterizations as available in the literature. Once the charmonium states are produced, some fraction of them may dissociate due to the inelastic collisions with the nucleons inside the nuclear medium they traverse. To account for this final state absorption of the charmonium states, we employ the Glauber model framework.  \\

In the leading-order (LO) calculation of perturbative QCD, the $c\bar c$ hadro-production cross section is the sum of two partonic level sub-processes: gluon fusion ($gg \rightarrow c\bar{c}$) and quark-antiquark annihilation ($q \bar{q}  \rightarrow c\bar{c}$), convoluted with the parton densities in the colliding hadrons. In p+A collisions, the LO $c\bar{c}$ production cross section is thus obtained by convoluting the sub-nucleonic cross sections  $\sigma_{gg \rightarrow c\bar{c}}$ and $\sigma_{q \bar q \rightarrow c\bar{c}}$ with the corresponding probabilities of finding the partons inside the projectile proton  and the target nucleus. \\


\begin{figure}
    \centering
    \includegraphics[width=1.0\linewidth]{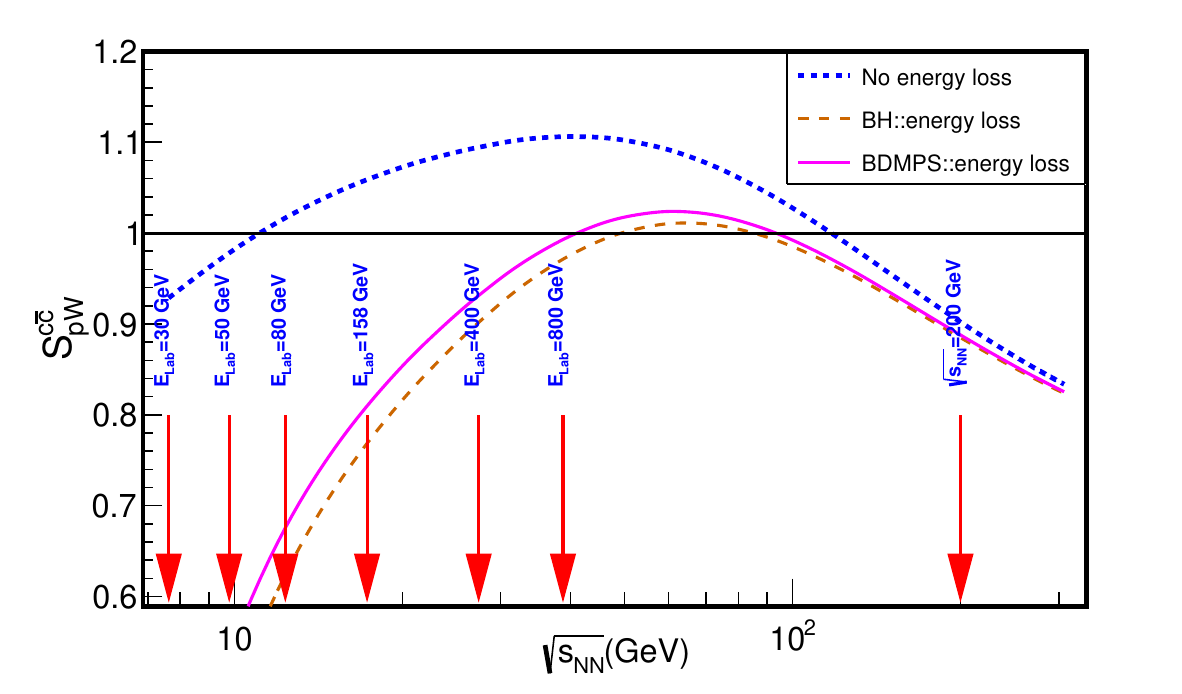}
    \caption{Variation of the shadowing ratio at mid-rapidity ($y_{cms}$ =0.0) as a function of centre-of-mass energy ($\sqrt{s_{\rm NN}}$) for p+W collisions. The blue (dot-dashed) line, magenta (solid dashed) line and the orange (dotted) line respectively denote evaluation of $c\bar{c}$ production cross section without any initial state energy loss and with BH and BDMPS schemes of beam parton energy loss. All the calculations are done using central sets of the CT18ANLO free proton PDF and EPPS21NLO nuclear PDF.}
    \label{fig:Sjpsi_vs_Elab}
\end{figure}


\par

Measurements like deep inelastic scattering (DIS) and Drell-Yan (DY) production using various nuclear targets have explicitly shown that the quark densities within the nuclei are significantly altered as compared to those inside the free protons. Such nuclear modifications depend on the parton's momentum fraction \( x \), the momentum scale ($Q^2$), and the target mass number, A. Both DIS and DY measurements directly probe the quark and antiquark distributions inside the nucleons. Gluons not being responsive to electromagnetic interaction, extraction of their distributions is an indirect process and relies on the scale-dependent structure function and the momentum sum rule for the quarks and gluons. Several groups, including DSZS~\cite{deFlorian:2011fp}, nCTEQ15~\cite{Kovarik:2015cma}, EPPS16~\cite{Eskola:2016oht}, and EPPS21~\cite{Eskola:2021nhw}, have recently conducted global fit extractions of the nuclear parton distributions (nPDFs). Additionally, other groups like MRSTW~\cite{Martin:2009iq} and CTEQ-TEA~\cite{Stump:2003yu} have extracted the free proton parton distribution functions (PDFs) across a wide kinematic range. For improved accuracy, CTEQ-TEA has recently incorporated a variety of new data from the LHC, including measurements of inclusive jet production, \( W \), \( Z \), and DY production, as well as the top quark pair production data from ATLAS, CMS, and LHCb, along with HERA (I + II) inclusive DIS data and jet production data from  FERMILAB . On the other hand, the EPPS16 nPDF scheme was the first to utilize the LHC data to more precisely constrain parton densities across a wider kinematic range relative to other existing methods. In comparison, the recently proposed EPPS21 nPDF routine incorporates additional data from p+Pb reactions at the LHC, such as the double differential di-jet production cross section measured by the CMS collaboration and D meson production by the LHCb collaboration at 5 TeV along with 8 TeV CMS data on production of \( W^{\pm} \) bosons. Furthermore, DIS measurements reported by the Jefferson Lab, that probe nuclear PDFs at both large and small virtualities, have also been included in the EPPS21 analysis.  Assuming factorization, the nuclear modification of parton densities with respect to  free protons  are parametrized as:

\begin{eqnarray}
f_{i}^{A}(x,Q^{2}) = R_{i}^{A}(x,Q^{2})f_{i}^{p}(x,Q^2)
  \label{Eq:RiA}
\end{eqnarray}
where the shadowing parameter, $R_{i}^{A}(x,Q^{2})$, converts the free proton distribution of the i-th parton into the bound one. In principle, shadowing parameters should also be dependent on the spatial position of the nucleons inside the target, with the nucleons at the core being more shadowed compared to those residing at the surface. In literature, such spatially inhomogeneous local shadowing effects on charm production in heavy ion collisions was first introduced in~\cite{Emelyanov:1997guf}. Subsequently, this was investigated in~\cite{Emelyanov:1998yul,Emelyanov:1999pkc,Klein:2003dj}, to study the effect of the impact parameter dependent nuclear parton densities on heavy quark and quarkonium production in nuclear collisions. The impact parameter dependence of the EPS09 and EKS98 nPDF schemes were also parameterized~\cite{Helenius:2012wd} following the specific target mass dependence of the EPS09 and EKS98 parameter sets. To produce the mass number independent coefficients, terms up to fourth order in the nuclear thickness was found to be necessary. The impact parameter dependence of the gluon shadowing was also addressed in~\cite{McGlinchey:2012bp} by analyzing the centrality and rapidity dependence of $J/\psi$ production measured in $\sqrt{s_{NN}} = 200$ GeV d+Au collisions at RHIC. The onset of nuclear shadowing was found to be a highly nonlinear function of collision centrality. In the present analysis, since we work with the p+A data samples integrated over collision centrality, the local shadowing effects are ignored. In our current study, we have thus employed the global EPPS21 NLO nPDF~\cite{Eskola:2021nhw} scheme, representing parton densities averaged over entire nucleus along with the CT18ANLO~\cite{Hou:2019qau} free proton PDF scheme  which is a extended family of CTEQ-TEA to model the parton densities inside the target and projectile respectively. In Fig.~\ref{Fig: pdf}, we have shown the variation of the shadowing parameter, $R_{i}^{W}(x,Q^{2})$, for the valence quarks and gluons, as a function of the target momentum fraction, $x_{2}$, evaluated at a momentum scale $Q = 3.1$ GeV, inside a tungsten (W) nucleus. For a given value of $x_{2}$, $R_{i} < 1$ indicates the depletion of the density of the i-th parton inside the nuclear target, whereas $R_{i} > 1$ denotes the enhanced density of bound partons. As evident from the figure, the degree of modification of parton densities is different for the valence quarks and gluons. \\

\begin{figure*}
    \centering
    \includegraphics[width=1.0\linewidth]{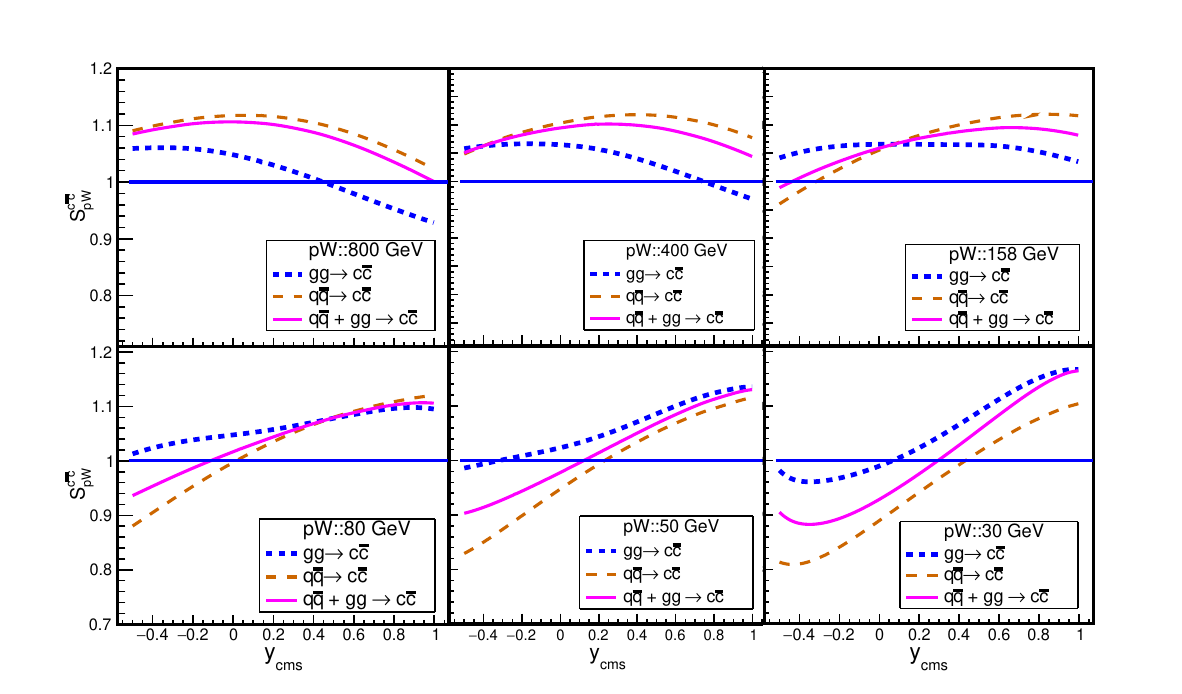}
    \caption{ 
        Variation of the shadowing ratio (magenta solid line) for p+W collisions, as a function of centre-of-mass rapidity $(y_{cms})$, at different beam energies relevant for existing or upcoming fixed target experiments, in absence of any initial state parton energy loss effect. The blue dotted line and orange dashed line represent the shadowing ratios due to individual processes, $gg \rightarrow c\bar{c}$ fusion and $q\bar{q} \rightarrow c\bar{c}$ annihilation, respectively. All the calculations are performed using central sets of the CT18ANLO free proton PDF and EPPS21NLO nuclear PDF.
       } 
    \label{fig:S_vs_xF}

\end{figure*}

Projectile partons, before undergoing hard collision with the target partons, may lose some energy during their passage through cold QCD matter of the target. Nowadays, several measurements at the RHIC and LHC, including the high momentum photon jet measurements and the DY measurements, have reaffirmed that partons lose energy due to multiple scattering with the surrounding QCD medium or through gluon radiation (i.e., QCD bremsstrahlung) while traversing the nuclear matter~\cite{Bjorken:1982tu,Cunqueiro:2021wls}. This energy loss also occurs for incident partons in the fixed target experiments and would modify the charmonium production in p+A collisions. Let a quark (gluon) from the projectile proton of energy $E_{b}$, lose $\Delta E_{q(g)}$ energy along its path as it moves through the nuclear medium before annihilating (fusing) with another anti-quark (gluon) inside the target. Consequently, the total shift in the momentum fraction ($\Delta x_{1}$) of this quark (or gluon) is given by
\begin{equation}
    \Delta x_{1q(g)}=\frac{\Delta E_{q(g)}}{E_b}
\end{equation}
We consider the gluon energy loss \((\Delta E_g)\) to be \(\frac{9}{4}\) times the quark energy loss \((\Delta E_q)\). This ratio is based on the color factors for the gluon-gluon interactions, which is 3, and for the quark-gluon interactions, which is \(\frac{3}{4}\). In the literature, different parametrizations have been independently proposed for modeling the fractional energy loss of projectile quarks inside the nuclear matter of the target. Within the Brodsky-Hoyer (BH) formalism~\cite{Brodsky:1992nq}, utilizing an analogy to the photon Bremsstrahlung process in quantum electrodynamics\linebreak (QED),  a formula for gluon radiation that accounts for the initial state quark energy loss is derived as:
\begin{equation}
    \Delta x_{1q} \approx \frac{\alpha}{E_{b}} <L>_{A}
\end{equation}
where $\alpha$ represents the specific energy loss of the incident quark in nuclear matter . The mean path length that the incident quark travels inside the target nucleus is given by \( <L>_{A} = \frac{3}{4}R_{A} \), where \( R_{A} \) is the radius of the nucleus, defined as \( R_{0}A^{1/3} \), assuming a uniform matter density distribution. It is important to note that the relevant length scale depends on the underlying process being computed. For initial state energy loss, as presently being studied, the average path length is calculated between the front surface of the nucleus to the point of hard interaction and thus appears to be $\frac{3}{4}R_{A}$. The same is true for any final-state effect, operative between the interaction point to the back surface of the nucleus. On the other hand, for the coherent energy loss scenario, amplitudes of the radiated gluons prior to and post hard interactions interfere and the relevant path length between front and back surface of the nucleus becomes $\frac{3}{2}R_{A}$. \\

On the other hand, the BDMPS~\cite{Baier:1996sk} approach, as introduced by Baier {\it et. al.} in 1996, expands upon the BDM model. Within this framework, the energy loss of sufficiently energetic partons is believed to depend on both the characteristic length and the broadening of transverse momentum (\( p_{T} \)) of the parton. For a finite-sized nucleus, both of these factors vary as \( A^{1/3} \), allowing us to quantify the mean energy loss as follows:
\begin{equation}
    \Delta x_{1q} \approx  \frac{\beta}{E_{b}} <L>_{A}^{2}
    \label{eloss_BDMPS}
\end{equation}
One may note that the BDMPS formalism examines the effects of initial state radiation in the Landau-Pomeranchuk-Migdal (LPM) regime, which occurs when formation time scales of the emitted gluons are comparable to the length of the medium. The original formulation of the BDMPS approach incorporates fluctuation in the induced energy loss, leading to an energy spectrum \(D(\epsilon)\) of the radiated gluons. In the current analysis, quenching is approximated by shifting the energy of the incident quark by a constant mean value, denoted as \(\Delta x_{1}\). The average energy loss of the incident quarks has a linear dependence on the traversed path length in the BH approach, while it has a quadratic dependency in the BDMPS framework. In a recent study~\cite{Giri:2025bfq}, we conducted a detailed systematic analysis of the DY reactions using these two initial state energy loss parameterizations. We determined the values of the parameters, $\alpha = 0.467 \pm 0.053 $ $GeV/fm$ and $\beta =0.068 \pm 0.008 $ $GeV/ fm^{2}$, for the incident quarks, by examining the available nuclear DY data on the differential production cross-section ratio. Values so obtained are now used to model the effect of initial stage  parton energy loss in $c\bar{c}$ production. As previously mentioned, the corresponding energy loss values for the incident gluons are obtained by scaling up the qurak energy loss by a factor of 9/4. Before moving forward, it might be interesting to note that other models of initial state quark energy loss are also prescribed in the literature. Within the Gavin and Milana (GM)~\cite{Gavin:1991qk} scheme, the initial state quark energy loss is assumed to grow linearly with the incident quark momentum fraction. As investigated in ~\cite{Song:2012zz}, the available experimental data from ${\pi}^{-} + A$ collisions at lower beam energies rule out the GM model. We thus refrain from using the GM energy loss scheme in the previous work as well as in the present study. The NVZ~\cite{Neufeld:2010dz} scheme as proposed by Vitev and collaborators, assumes a linear path length dependence of the initial state quark energy loss in cold nuclear matter, similar to the BH picture. Hence we do not consider this model separately in our present study. \\

Considering both incoming quark and gluon energy loss inside target nucleus, the momentum fraction of the incident parton will undergo a shift from \(x_{1}^{'} = x_{1} + \Delta{x_{1}}\) to \(x_{1}\) at the point of fusion. Thus, to calculate the $c\bar{c}$ production cross-sections, the parton density function, \(f_i^p(x, Q^2)\), of the projectile proton must be evaluated at \((x_{1} + \Delta{x_{1}})\). Due to the steep nature of the gluon and valence quark distributions at large \(x_{1}\), even a small shift \(\Delta{x_{1}}\) can lead to substantial suppression of the charmonium production cross-section in a heavier nucleus compared to a lighter one. Taking into account both initial state nuclear shadowing and beam parton energy loss and no final state dissociation, the single differential production cross-section for the i-th charmonium state, in p+A collisions, can be written as:
\begin{eqnarray}
     &\frac{d\sigma^{i,0}_{pA}}{dy_{cms}}=F_i\int_{2m_c}^{2m_D} \frac{2mdm}{s} \Bigg\{f_g^p(x_1',m^2)f_g^A(x_2,m^2)\sigma_{gg}(m^2)\nonumber \\ &+ \sum_{q=u,d,s}\Big{[}f_q^p(x_1',m^2)f_{\bar{q}}^A(x_2,m^2)\nonumber \\&+f_{\bar{q}}^p(x_1',m^2)f_q^A(x_2,m^2)\Big{]}\sigma_{q\bar{q}}(m^2)\Bigg\}
    \label{Eq:CEM4}
\end{eqnarray}

where $x_{1(2)}$ is the Bjorken-$x$ variable and specifies the fraction of the projectile(target) momentum carried by a parton. $f(x,m^{2})$ is parton density probed at a momentum scale $Q^{2} \equiv m^{2}$ with momentum fraction $x$. For such $2 \rightarrow 1$ process, the Feynman scaling variable $x_F( \equiv  x_1- x_2)$ is related to the $c\bar{c}$ pair centre of mass rapidity ($y_{cms}$) via  $x_{F} = {2m \over \sqrt{s_{\rm NN}}} sinh (y_{cms})$, where $\sqrt{s_{NN}}$ is the energy of the colliding hadrons in the centre of mass frame. In the CEM model, since all charmonium states are treated identically, it provides a comprehensive energy and momentum distribution of the various states. To determine the constant fraction $F_i$, the transition probability to each state is required. It is important to note that $F_{J/\psi}$ accounts for both direct $J/\psi$ production and contributions from the higher excited states, such as $\chi_{cj}$ through radiative decay and $\psi(2S)$ through hadronic decay and fixed from the data. \\

\begin{figure*}
    \centering
    \includegraphics[width=1.0\linewidth]{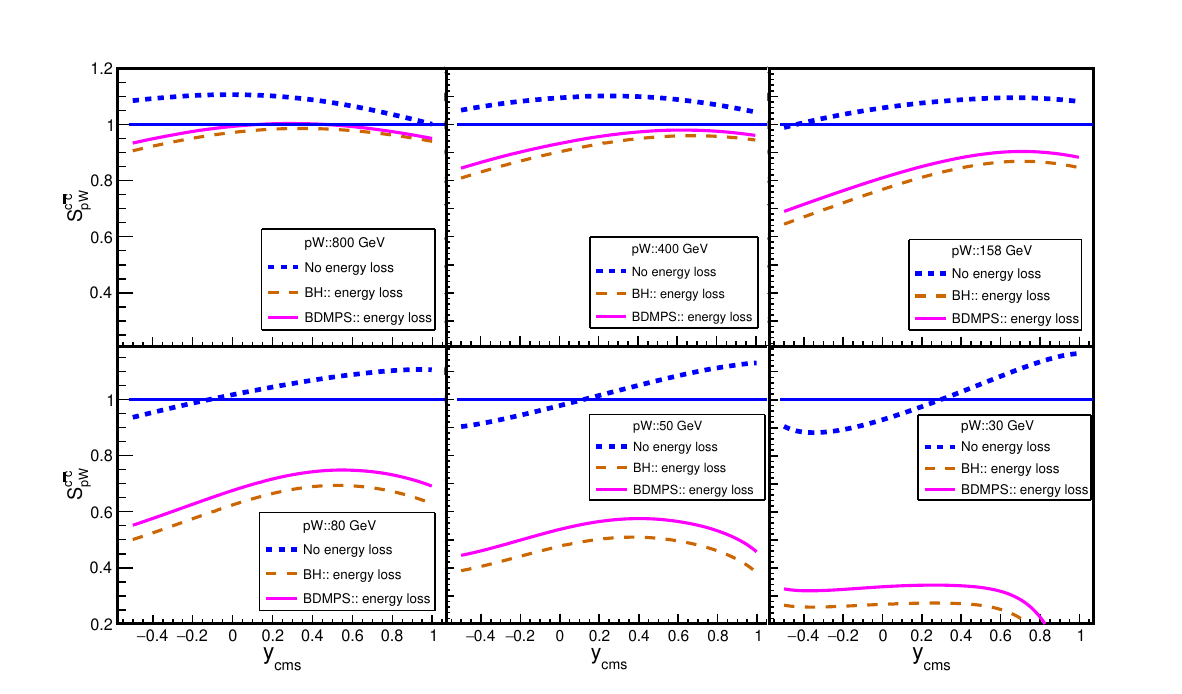}
 
        \caption{
        Variation of the shadowing ratio for p+W collisions, as a function of centre-of-mass rapidity $(y_{cms})$, relevant for existing or upcoming fixed target experiments after accounting for energy loss of projectile partons prior to hard interactions. The dotted (blue) line corresponding to no initial state energy loss scenario, the dashed (orange) line corresponds to the linear path length dependence of energy loss following BH scheme and the solid (magenta) line corresponds to quadratic path length dependence of energy loss following BDMPS scheme. All the calculations are carried out with the CT18ANLO free proton PDF and EPPS21NLO nuclear PDF.
        }
    \label{fig:S_vs_xF_DiffModel}
\end{figure*}


At this juncture, it would be interesting to quantitatively assess the influence of the employed initial state effects on the perturbative charm production cross section in p+A collisions. For this purpose, let us now calculate the shadowing ratio, $S_{pA}^{c\bar{c}} (y_{cms})$, defined by the ratio between the perturbative per nucleon $c\bar{c}$ production cross sections in p+A and p+p collisions as:

\begin{equation}
    S^{c\bar{c}}_{pA}(y_{cms})=\frac{1}{A}\frac{d\sigma^{c\bar{c}}_{pA}/dy_{cms}}{d\sigma^{c\bar{c}}_{pp}/dy_{cms}}
    \label{Eq:shad_ratio}
\end{equation}
\par
Shadowing ratio for charm production has  previously been  calculated in Refs.~\cite{Lourenco:2008sk, Arnaldi:2009it, Bhaduri:2014daa}.  In all those calculations $S^{c\bar{c}}_{pA}(y_{cms})$ was only sensitive to the underlying nuclear parton distribution functions. However, in our present study the presence of energy loss of projectile partons is likely to influence the results. 
Since tungsten has been commonly used in all the fixed target p+A collisions studying $J/\psi$ production, we first estimate the variation of the mid rapidity ($y_{cms}=0.0$) shadowing ratio in p+W collisions, as a function of collision energy. 
Before we proceed further, it may be noted that in the present study (as well as in the previous study on DY production reported in~\cite{Giri:2025bfq}), we have convoluted the NLO PDFs with LO partonic cross sections to estimate $c\bar{c}$ production in p+A and p+p collisions. The EPPS21 (or even EPPS16) nPDF routines are not provided for LO sets. Using NLO PDFs with LO calculation is not fully consistent and can only be accepted as an approximation where one neglects the higher-order terms in the QCD coefficient functions, while keeping the NLO evolution. However the resulting effect of this mismatch between LO calculation with NLO nPDF and free proton pdf on the initial state energy loss or final state absorption, in case of $J/\psi$ production, is found to be small and will be discussed later, in detail. \\

Results of our calculations of the shadowing ratio are displayed in Fig.~\ref{fig:Sjpsi_vs_Elab}. In absence of any initial state parton energy loss, $S^{c\bar{c}}_{pW}$ is greater than unity indicating an enhanced $c\bar{c}$ production in p+W collisions in the nucleon-nucleon centre of mass energy range, $\sqrt{s_{NN}}$, between 13 - 100 GeV (equivalently beam energy ($E_{b}$) range between 80 - 5000 GeV), with a broad maximum around $\sqrt{s_{NN}} \simeq 45$ GeV ($E_{b} \simeq$ 1000 GeV). 
The corresponding $x_{2}$ values range  from 0.03-0.25, where both quark and gluon distributions bound inside W nucleus exhibit an enhancement compared to free protons, as evident from Fig.~\ref{Fig: pdf}. An overall shadowing effect (ratio less than unity) is observed around  the beam energy 50 GeV and below. Accounting for the energy loss  of beam partons, prior to hard interactions, significantly influences the shape of the shadowing ratio as a function of $\sqrt{s_{\rm NN}}$. The anti-shadowing behavior, in absence of any energy loss effect, was first observed in Ref.~\cite{Lourenco:2008sk}. In presence of initial state energy loss, this turns into a shadowing behavior, with BH energy loss scheme generating  a little less  suppression than the BDMPS scheme. The difference in the $c\bar{c}$ production cross sections increases with lowering the energy of the collisions. Beyond $\sqrt{s_{\rm NN}} > 100$ GeV the two energy loss models produce almost identical suppression of charm production in p+W collisions, which is also similar in shape and magnitude to the shadowing effects exhibited by the depletion in the target parton densities alone for the no energy loss scenario. the role of the initial state parton energy loss on charm production in p+A collisions thus appears not to be very significant at collider energies. \\

\begin{figure*}
    \centering
    \includegraphics[width=1.0\linewidth]{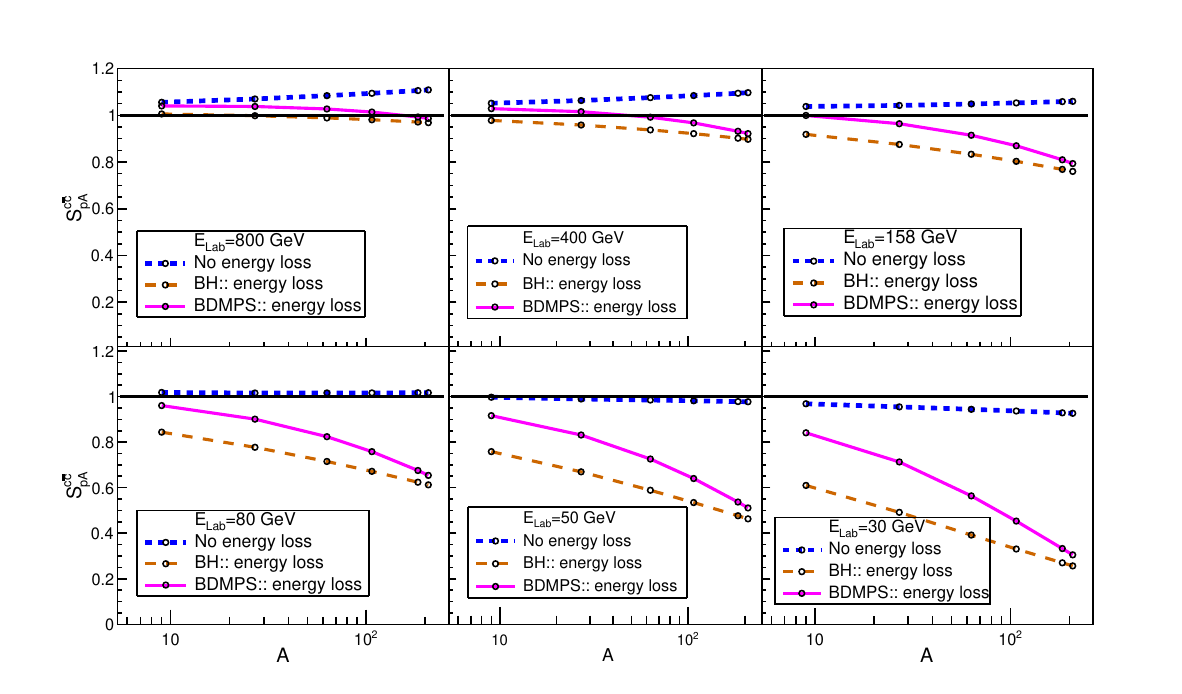}
    \caption{
        Variation of the mid-rapidity shadowing ratio as a function of target mass (A) in fixed target p+A collisions with incident proton beam energies ranging between 30 - 800 GeV relevant for available and upcoming measurements. The dotted (blue) line corresponding to no initial state energy loss scenario and incorporates only the effect of modified parton densities in the target. The dashed (orange) line corresponds to the linear path length dependence of energy loss following BH scheme and the continuous (magenta) line corresponds to quadratic path length dependence of energy loss following BDMPS scheme. All the estimations are evaluated with the CT18ANLO free proton PDF and EPPS21NLO nuclear PDF.
       }
    \label{fig:S_vs_L}
 \end{figure*}

With an aim to study the nature of the energy and kinematic dependence of the shadowing ratio in further details, we now calculate the variation of $S^{c\bar{c}}_{pW}$ with $y_{cms}$, at different beam energies relevant for the existing or the upcoming fixed target measurements as respectively shown in Fig.~\ref{fig:S_vs_xF} (Fig.~\ref{fig:S_vs_xF_DiffModel}) in absence(presence) of the projectile parton energy loss scenarios. Since no data exist or is foreseen at very high negative rapidities, we limit our calculation till $y_{cms} = -0.5$ in the backward hemisphere. For the no energy loss scenario, contributions from each of the two LO partonic sub-processes are also shown separately, to understand their relative influence on the overall shadowing ratio. As evident, the contribution due to $gg$ fusion dominates at 800 GeV but as the energy decreases, there is a transition from $gg$ dominance of the shadowing ratio to $q\bar{q}$ dominance.
The observed feature can be attributed to the dominance of the valence quark densities over the gluons with decreasing centre-of-mass energy of the collision ($\sqrt{s_{NN}}$), leading to increase in $x_1$, for a given $y_{cms}.$
Once the energy loss effect is turned on, the previously observed enhancement now leads to a depleted charm production over the entire rapidity space, with stronger reduction at the lower energies. The small difference between the two energy loss models become insignificant as we gradually increase the energy of the collision.\\
\begin{table*}
 \centering   
\begin{tabular}{|c|c|c|c|c|c|}
\cline{1-6}
  Experiment   & $E_{b}$(GeV) & Targets &data type& rapidity range & $x_{F}$ range  \\
  \cline{1-6}
  NA50   & 400 & Be,Al,Cu,Ag,W,Pb & Integrated&$-0.425<y_{cms}<0.575$ &  $-0.099<x_{F}<0.137$\\
  \cline{1-6}
  NA50  &450 & Be,Al,Cu,Ag,W & Integrated&$-0.5<y_{cms}<0.5$& $-0.111<x_{F}<0.111$ \\
  \cline{1-6}
  NA60 &400& Be,Cu,In,W,Pb,U & Integrated&$-0.17<y_{cms}<0.33$& $-0.038<x_{F}<0.076$ \\
  \cline{1-6}
  NA60 &158 & Be,Al,Cu,In`,W,Pb,U & Integrated & $0.28<y_{cms}<0.78$& $0.101<x_{F}<0.308$\\
  \cline{1-6}
  HERA-B &920& C,W & Differential&$-1.56<y_{cms}<0.84$& $-0.34<x_{F}<0.14$ \\
  \cline{1-6}
  E866 &800& Be,Fe,W & Differential&$-0.6<y_{cms}<2.46$& $-0.10<x_{F}<0.93$ \\
  \cline{1-6}
  E906 &120 & C,Fe,W & Differential & $0.86<y_{cms}<1.47$& $0.4<x_{F}<0.85$\\
  \cline{1-6}
\end{tabular}
\caption{Basic features of the selected fixed target  experiments measuring charmonium production in p+A collisions.}
\label{tab:Expt. details}
\end{table*}

Finally, to investigate the target mass dependence, we plot in Fig.~\ref{fig:S_vs_L} the mid-rapidity shadowing ratio for various light to heavy target nuclei, in fixed target p+A collisions, at different incoming proton beam energies between 30 - 800 GeV. As expected at higher beam energies, in the absence of beam parton energy loss, the ratio is greater than unity, indicating an enhancement in the $c\bar{c}$ production and hence an anti-shadowing effect, the magnitude of which increases for heavier targets. With lowering the beam energy this eventually gets converted into suppressed $c\bar{c}$ production or a shadowing effect below 80 GeV solely due to the depleted parton densities (more depletion for heavier masses) inside the target. Once we consider the energy loss of the beam partons, an overall shadowing is observed, which grows stronger in magnitude with increasing target mass and more importantly with decreasing beam energy. As evident from the figure, at lower energies the influence of energy loss phenomenon is much stronger as compared to nuclear modification of parton densities inside the target. The difference in the resulting suppression patterns between the two models of energy loss also  grows with lowering energy of the collision, with the BH model generating a stronger reduction in the $c\bar{c}$ production cross section as compared to the BDMPS scheme.  \\



\begin{table}[htpb]
    \centering
    \begin{tabular}{|c|c|c|c|}
        \cline{1-4}
        Experiment & Energy loss & $\sigma^{J/\psi}_{abs}(mb)$ & $\chi^2/ndf$ \\ 
        & model& &  \\ \cline{1-4}
            \multirow{3}{*}{NA50 $\&$ NA60 400 GeV}  & No   &  4.91$\pm$0.5  & 0.47   \\ 
        \cline{2-4}
          & BH   & 2.50$\pm$0.6   & 0.51   \\ 
          \cline{2-4}
           & BDMPS  & 2.40$\pm$1.0  & 0.63  \\ \cline{1-4}

           \multirow{3}{*}{\shortstack{NA50  450 GeV \\ (LI $\&$ HI)}} & No  & 4.56$\pm$1.0  & 0.46  \\ \cline{2-4}
          & BH  & 2.11$\pm$0.6   & 0.37  \\ \cline{2-4}
          & BDMPS  & 2.03$\pm$0.6  & 0.31  \\ \cline{1-4}
       
           \multirow{3}{*}{NA60 158 GeV}  & No  & 7.31$\pm$0.6&0.15  \\ \cline{2-4}
          & BH  & 3.84$\pm$0.5  & 0.91  \\ \cline{2-4}
          &BDMPS &3.46$\pm$0.5  & 0.68\\ \cline{1-4}
    \end{tabular}
      \caption{
        Extracted final state absorption cross section of $J/\psi$ mesons ($\sigma^{J/\psi}_{abs}$) obtained from the fit to the ratio of the per nucleon $J/\psi$ production cross sections measured by the NA50 and NA60 experiments in fixed target $p+A$ collisions in the SPS energy domain, under various scenarios of initial state parton energy loss. Combined fits are performed for the NA50 HI and LI data sets at 450 GeV and for the NA50 and the NA60 data sets at 400 GeV}.
        
    \label{tab:Fitting result}
\end{table}

Once produced, the nascent $c\bar{c}$ pairs undergo final state interaction in the nuclear medium till they exit the target nucleus. Within the nuclear absorption scenario, as being adopted in this work, the dominating final state interaction is the inelastic collision of the target nucleons with the correlated color neutral $c\bar{c}$ pairs in their pre-resonance or resonance stage, with an absorption cross section $\sigma_{abs}$ quantifying the final state dissociation. 
A widely used prescription to treat the charmonium absorption in the nuclear medium is realized within the framework of the Glauber model~\cite{Kharzeev:1996yx}. 
The key components for the Glauber formalism include the nuclear density profiles, $\rho_A$, and the interaction cross-section for the reaction of interest. Using a Wood-Saxon nuclear density distribution profile with specific parameters provided in~\cite{DeJager:1974liz,DeVries:1987atn}, we obtain the per nucleon charmonium production cross section in p+A collisions as:
\begin{eqnarray}
    \frac{\sigma^{i}_{pA}}{A}=\frac{\sigma^{i,0}_{pA}}{(A-1)\sigma^{i}_{abs}}\int \vec {db}(1-exp(-(A-1)\sigma^{i}_{abs}T_A(\vec{b})))
    \label{Eq:Full Glauber}
\end{eqnarray}
where $\sigma^{i,0}_{pA}$ is the $i^{th}$ resonance state production cross section in $p+A$ collisions where no nuclear absorption takes place, $\sigma^{i}_{abs}$ is the final state absorption (or dissociation) cross section of $i^{th}$ charmonia state in the nuclear medium and $T_{A}(~b)$ is the nuclear thickness function, representing the number of nucleons
per unit surface at an impact parameter $\vec{b}$ of target of mass number $A$.

\par

\section{Selection of experimental data}

Several fixed-target experiments, measuring charmonium production in p+A collisions, have been conducted at the accelerator facilities like CERN SPS, Fermilab Tevatron and DESY. A summary of the main features of these experiments, we plan to examine for the present study, including their beam energy, exposed targets and phase space coverage is given in Table~\ref{tab:Expt. details}. Note that we have not included some of the old measurements in our table. Before E866 experiment, Ferimilab E772 experiment also recorded data on the production of $J/\psi$  and $\psi(2S)$ states with 800 GeV proton beam incident upon deuterium (d), carbon (C), calcium (Ca), iron (Fe), and tungsten (W) targets. E866 experiment later collected data at the  same beam energy but with much wider kinematic coverage. Also subsequent analysis of the E772 data set within the E866 analysis framework revealed that the E772 results were biased due to wrongly corrected detector acceptance. Similarly, the NA38 measurements at the SPS, later evaluated with the NA50 muon analysis software, were found to be affected by incorrectly evaluated reconstruction efficiencies. the NA51 experiment published data only for p+p and p+d collisions, and hence is not suitable for studying CNM effects. Results from these old fixed target experiments are thus excluded from our list. Among our listed experiments, the E866 experiment bombarded beryllium (Be), iron (Fe) and tungsten (W) targets with an 800 GeV proton beam and accumulated the differential cross section of different charmonia states over a rather broad kinematic regime. At the CERN SPS, the NA50 experiment carried out a detailed  measurement of the inclusive $J/\psi$ and $\psi(2S)$ production in p+A collisions, using  proton beam of energy 400(450) GeV, made incident on six (five) different targets (Be, Al, Cu, Ag, W, Pb). The most recent measurement from the CERN SPS facility is reported by the NA60 experiment that recorded $J/\psi$ production in 400 GeV and 158 GeV p+A collisions by using seven different nuclear targets. The most recent charmonium measurement in the low energy domain is from the SeaQuest/E906 Collaboration at the Fermilab. In this experiment, results are reported in terms of W/C and Fe/C differential $J/\psi$ production cross section ratios as a function of $x_{F}$ as well as transverse momentum $p_{T}$, in the forward $x_{F}$  regime. \\

To decide the suitability of the available data sets to be in accord with the nuclear absorption scenario, we calculate the $J/\psi$ formation length in the target rest frame corresponding to each experimental measurement in the probed kinematic domain. If $\tau_{0}$ denotes the intrinsic resonance formation time in the charmonia rest frame, the corresponding formation length$(d_0)$ in the target nucleus rest frame is given by:

\begin{eqnarray}
\centering
\hspace{2cm}
   d_{0} = \tau_{0} sinhy   
\end{eqnarray}
where $y=0.5 \times ln{{E+p_{z}} \over {E-p_{z}}}$ denotes the rapidity of the $c\bar{c}$ state in the target nucleus rest frame. As we have adopted the LO pQCD framework, $c\bar{c}$ pairs are produced with zero transverse momentum ($p_T =0$) and with energy, $E = \sqrt{M^2+p_z^2}$, where $M$ denotes the mass and $p_{z}$ denotes the longitudinal momentum of the pair in the target nucleus rest frame.
Following the additive property of the rapidity variable under Lorentz transformation, the resonance rapidity in the target rest frame can be calculated from the corresponding centre-of-mass rapidity, $y_{cms}$ (or $x_F$), as given in Table~\ref{tab:Expt. details}.
If $d_{0}$ is smaller than $ 5$ fm, the typical length of the nuclear matter traversed by the $c\bar{c}$ inside a heavy nuclei ($A \sim 200$), color neutralization and/or resonance formation occurs inside the target nucleus and nuclear absorption of these resonating or resonant $c\bar{c}$ pairs become the dominant final state interaction. On the other hand, if $d_{0}$ is substantially large compared to the average length of the nuclear matter, the nascent $c\bar{c}$ pairs remains colored over the entire crossing time inside the target. Color neutralization and resonance formation of such hard $c\bar{c}$ pairs occur outside the nuclear medium and dissociation of such pairs due to inelastic collision inside the target would be negligible. 

\begin{figure*}[htpb]
\centering
    \includegraphics[scale=0.90]{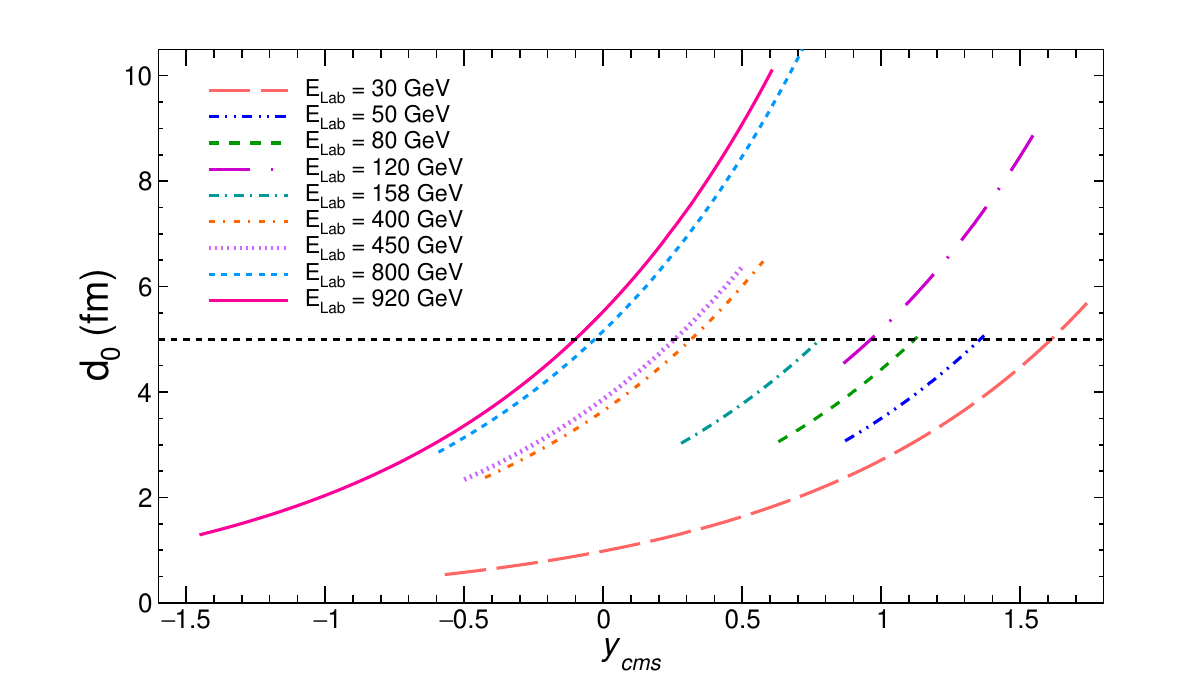}
    \caption{Variation of the $J/\psi$ formation length ($d_0$) in the nuclear rest frame as a function of centre-of-mass rapidity ($y_{cms}$) in the kinematic domain specific to each fixed target p+A collision experiment as listed in Table~\ref{tab:Expt. details}. In addition, the variation of $d_{0}$, in the foreseen kinematic window of the upcoming NA60+, J-PARC and CBM experiments are also shown. The $y_{cms}$ coverages at beam energies of 50 GeV and 80 GeV are calculated assuming the same laboratory frame rapidity ($y_{Lab}$) coverage of the NA60 and the NA60+ di-muon spectrometers. At the beam energy of 30 GeV, the corresponding $y_{cms}$ coverage is calculated from the foreseen angular coverage of the CBM STS detector system. The straight line parallel to x-axis denotes the average path traversed by the $c\bar{c}$ pairs inside the nuclear medium for a typical heavy target with $A=200$.}
    \label{fig:Formation Length}
\end{figure*}

Till date, no unique prescription is available in the literature to determine the quarkonium formation time. In Ref.~\cite{Kharzeev:1995br} the intrinsic color neutralization time, defined as the time required by a color octet $c\bar{c}$ pair to become a color singlet state via evaporation of a soft gluon, is estimated to be $\tau_0 \simeq 0.25$ fm. Time required to form the physical resonances with required size and right quantum numbers is believed to be slightly larger. Studies based on the potential theory suggest an intrinsic resonance formation time, $\tau_R \simeq 0.3$ fm for $J/\psi$, a value in line with the model independent upper bound based on the uncertainty principle and the separation between the ground state and first excited state. Alternative estimations of $\tau_R$ for various charmonium states are also available in literature. Reconstruction of the quarkonium formation dynamics from the experimental data on $e^{+}e^{-} \rightarrow Q\bar{Q}$ annihilation, using dispersion relations in a model independent way, predicts a distribution of $\tau_{J/\psi}$ with a mean value of 0.44 fm and a spread of 0.31 fm. Recently an attempt has been made to extract the charmonium formation time by analyzing the measured charmonium production cross sections in p+A collisions, leading to relatively lower value, $\tau_R \approx 0.1$ fm~\cite{Ferreiro:2021kwk}. Since within our nuclear absorption scenario, there is no clear distinction between a resonating and a resonant $c\bar{c}$ pair, we choose $\tau_0 = 0.25$ fm in our calculations, as the resonance formation time. Figure~\ref{fig:Formation Length} illustrates the variation of the resonance formation length, as function of $y_{cms}$, at different beam energies, relevant for the available and upcoming fixed target experiments, in their respective kinematic domain (see Table~\ref{tab:Expt. details}). As far as the future measurements are concerned, the $y_{cms}$ coverage at beam energies of 50 GeV and 80 GeV are evaluated assuming same di-muon rapidity coverage of the NA60 and the NA60+ experiments in the laboratory frame. At 30 GeV, the $y_{cms}$ is calculated from the foreseen angular coverage ($2.5^{o} < \theta_{\mu\mu} < 25^{o}$) of the silicon tracking system (STS), the key detector of the CBM experiment~\cite{CBM:2025mnp}. As evident from the figure, the kinematic domains probed by charmonia data collected by the E866 and E906  experiments are mostly focused at forward rapidities and thus unsuitable to the  nuclear absorption scenario. The HERA-B experiment at DESY recorded differential charmonium production cross sections at backward $x_F$ which is fit to the final state absorption scenario. It is usual to introduce $x_F$ dependent absorption cross section$(\sigma_{abs}^{\psi_i}(x_F))$ to explain this kind of differential data, as discussed in~\cite{Duan:2010zz,Lourenco:2008sk}. However, for the present analysis, it is difficult to carry out a comprehensive phenomenological study of the kinematics dependent charmonium absorption with only one measurement of differential data on charmonium production, which is congruous to this particular absorption picture. Formation lengths probed by the existing SPS data are small enough to ensure the color neutralization inside the nuclear medium making nuclear absorption as a favorable scenario to model the final state interaction. Hence for further analysis we choose the inclusive $J/\psi$ production cross sections on various targets published by the NA50 and the NA60 experimental collaborations at the  SPS facility. Our selection of analyzing only SPS data within the nuclear absorption scenario remains unaffected even if we calculate the formation length using an intrinsic formation time, $\tau_{0} \sim 0.4$ fm. 

\begin{figure*}
\centering
    \includegraphics[width=0.49\linewidth]{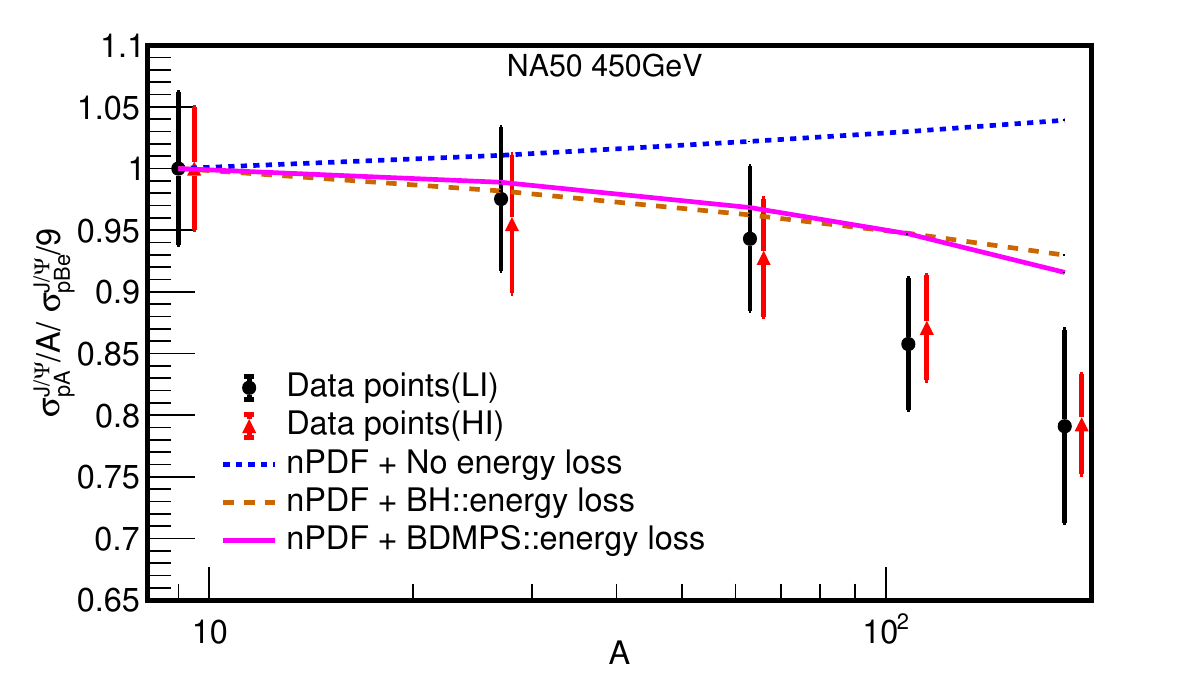}
     \includegraphics[width=0.49\linewidth]{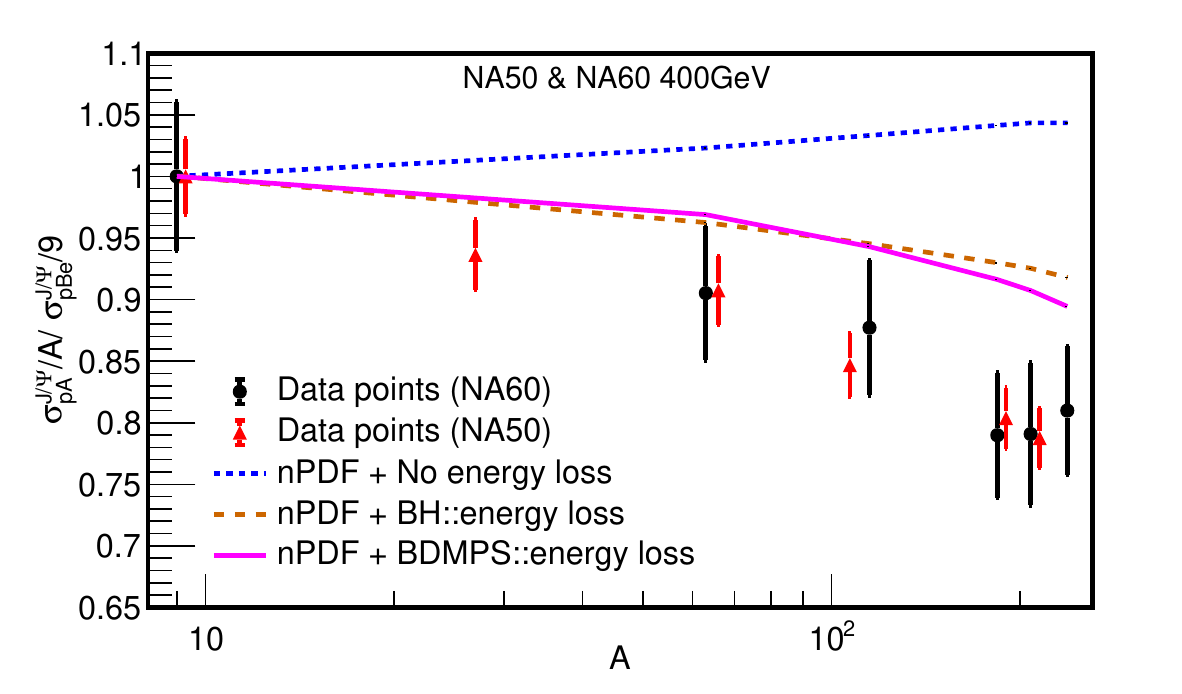}
     \includegraphics[width=0.49\linewidth]{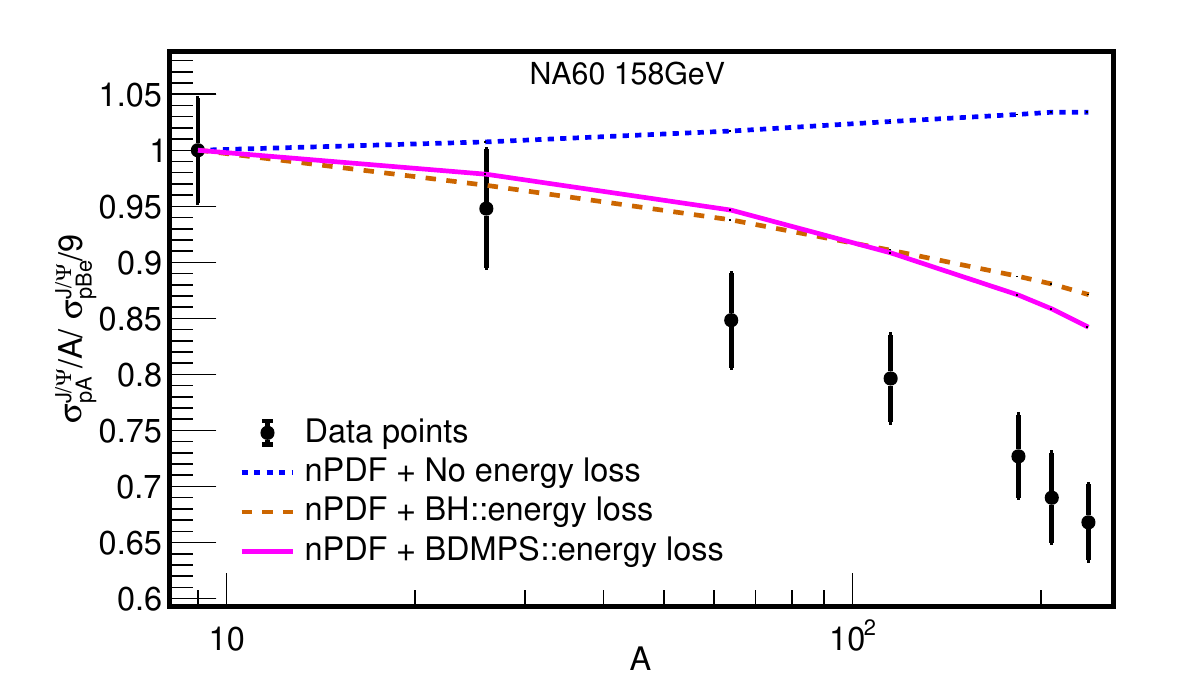}

     \centering
     \caption{Target mass ($A$) dependence of the ratio of the per nucleon $J/\psi$ production cross sections in p+A to p+Be collisions as measured at the SPS facility by the NA50 Collaboration at 450 GeV (top left), by the NA50 and NA60 Collaborations at 400 GeV (top right) and by the NA60 Collaboration at 158 GeV (bottom). At 450 GeV high intensity (HI) and low intensity (LI) data sets are separately shown in the same plot. At 400 GeV, the phase space coverage of the NA60 data points lies entirely within that of the NA50 data set.} The solid lines correspond to  our model calculations of the same ratio, for three scenarios of initial state CNM effects: modification of target parton densities (nPDF) in absence of any beam parton energy loss (blue dotted line), modification of target parton densities and the projectile parton energy loss inside the target following BH (orange dashed line) scheme and BDMPS (magenta continuous line) scheme. No final state CNM effect on $J/\psi$ production is taken into consideration.
     \label{fig: only Initial state effect}
\end{figure*}

\begin{figure*}
\centering
     \includegraphics[width=0.49\linewidth]{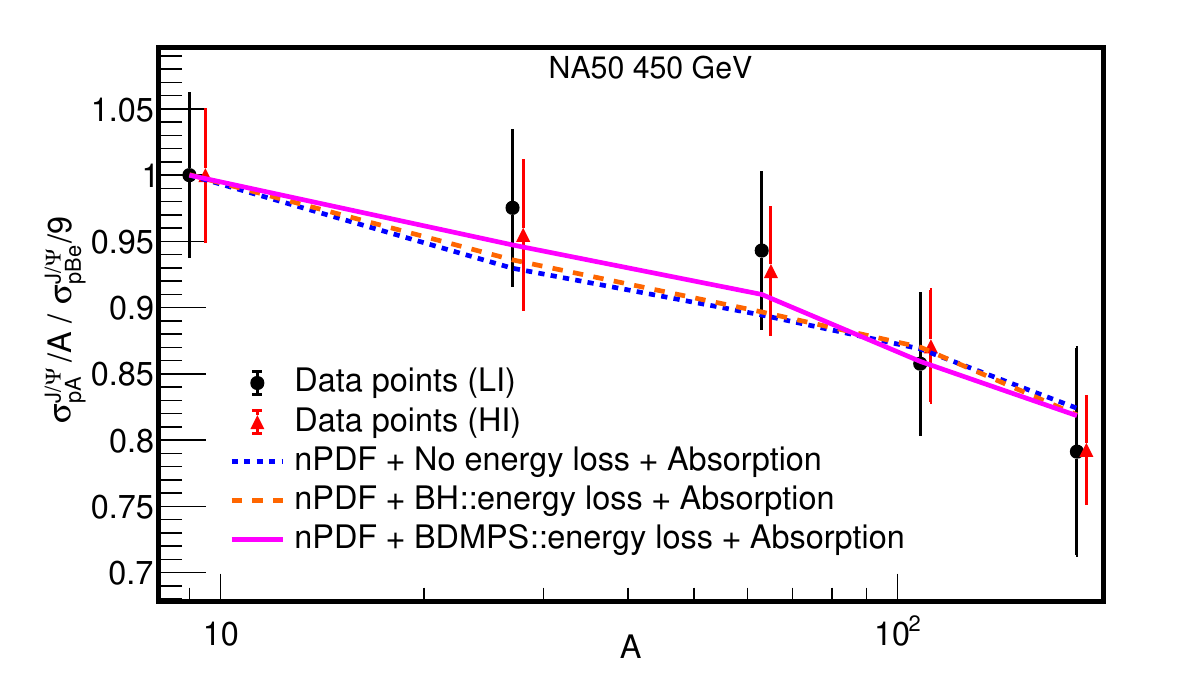}
     \includegraphics[width=0.49\linewidth]
     {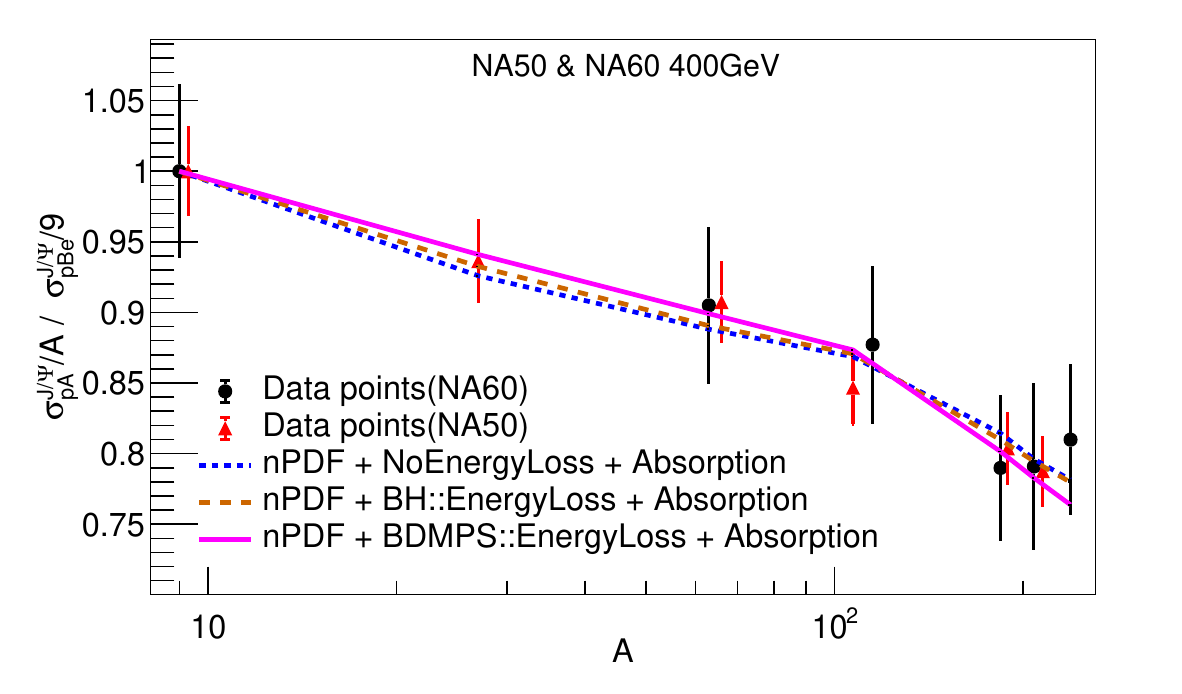}
     \includegraphics[width=0.49\linewidth]{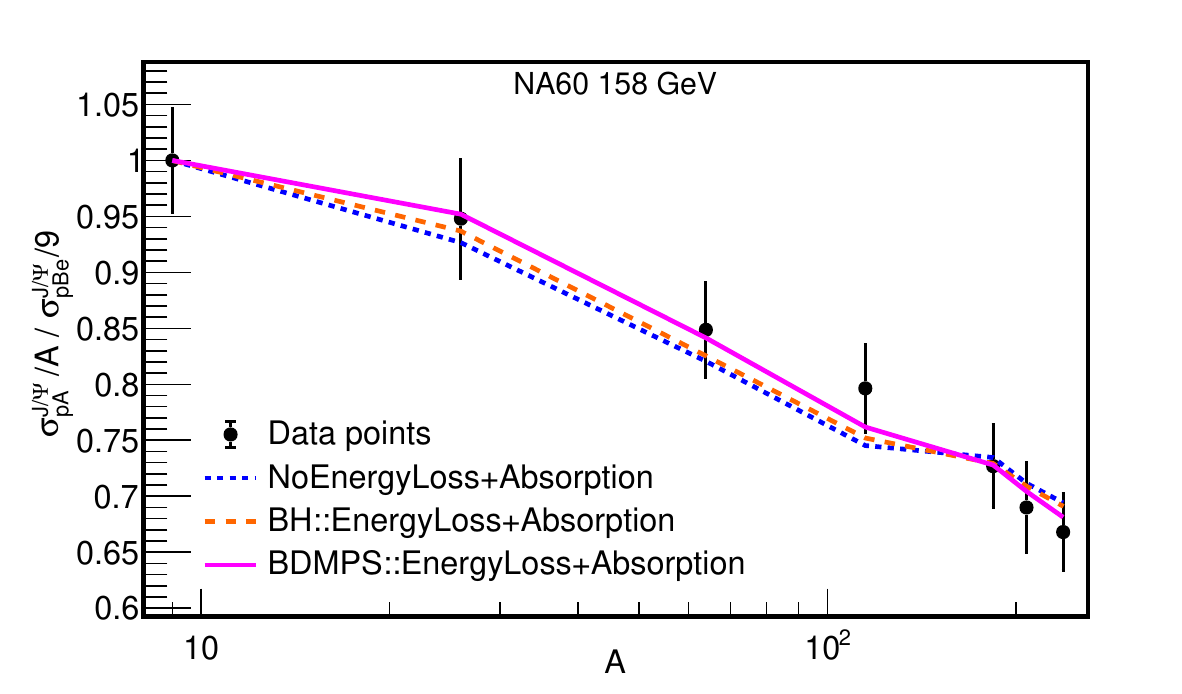}
     \caption{Ratio of the per nucleon $J/\psi$ production cross sections in p+A to p+Be collisions as a function of target mass ($A$) as measured at the SPS facility at 450 GeV (top left) by the NA50 experiment, at 400 GeV (top right) by the NA50 and NA60 experiments and at 158 GeV (bottom) by the NA60 experiment. The three different lines correspond to the fitting of our model calculations to the available data for three different theoretical scenarios: nuclear modification of parton densities (nPDF) plus the final state absorption(dotted line), nuclear modification of parton densities (nPDF) along with incident parton energy loss following BH (dashed line) and BDMPS (solid line) schemes in presence of final state absorption. At 450 GeV a combined fit to both high intensity (HI) and low intensity (LI) data sets is performed. Similarly at 400 GeV, a combined fit is performed to both NA50 and NA60 data sets. Though different, but the kinematic range of NA60 data corpus lies entirely within the range of NA50 data. The extracted $\sigma_{abs}^{J/\psi}$ values are listed in Table 2.}
    
     \label{fig:fitting plot}
 \end{figure*}

\section{Results and Discussion}



In this section, we discuss the extraction of the final state absorption cross sections from the available SPS data, its plausible dependence on the initial state effects and the energy of the collision. To better understand the role of the initial state energy loss, we first compare the data for $J/\psi$ production, with our model calculations, in absence of any final state dissociation, as displayed in Fig.~\ref{fig: only Initial state effect}. In accordance with the data published by the NA60 experiment, we plot the ratio of the per nucleon $J/\psi$ production cross section in p+A collisions to that in p+Be collisions. As all the targets are exposed to the incident proton beam in the same run, such ratios help to get a better control over the systematic uncertainties. Phenomenologically, such a ratio essentially helps us to get rid of the normalization parameter ($F_{i}$ in our case) associated with the model. Similar to our previous observations, we find that the presence of initial state parton energy loss, converts the anti-shadowing effects due to enhanced target parton densities into an effective shadowing, leading to reduced charm production in the SPS energy domain. Larger be target mass, stronger is the depletion, though the two models of energy loss generate comparable suppression. \footnote{In Fig.~\ref{fig: only Initial state effect}, the visible differences in the initial state energy loss between 158 and 400/450 GeV or the rise of the $J/\psi$ production cross section with target mass in the no energy loss scenario due to anti-shadowig effects in the SPS kinematic domain get obscure to some extent in the plotted ratio of the cross section to that of $A = 9$ (Be). In comparison to this ratio, nuclear modification factor, $R_{pA}$, is a better observable to investigate the above features.  As our intention is to understand the role of different CNM effects to reproduce the overall suppression observed in the available data and since NA50 and NA60 experiments have not measured $J/\psi$ production cross section in p+p collisions hence we refrain from plotting the observable $R_{pA}$. The reader may also note the fact that when final state absorption is not considered, the mid-rapidity shadowing ratio plotted in Fig.~\ref{fig:S_vs_L} is equivalent to $R_{pA}$ within the CEM framework (apart from not being integrated over the specific rapidity window) and displays the features stated above.} The model calculations clearly underestimate the observed suppression pattern present in the data, leaving room for additional dissociation mechanisms to become operative. As observed in~\cite{Vogt:1999dw}, an issue  with these energy dependent energy loss models is that at a beam energy of 800 GeV, $\Delta{x_{1}}$ grows larger than $x_{1}$ in the backward rapidity ($x_F < 0)$ region. We have not encountered any such problem in the probed kinematic region, down to the lowest beam energy investigated in this work. \\

We now move on to extract $\sigma_{abs}$ from each of the selected data sets by fitting it with Eq.~\ref{Eq:Full Glauber}, in the corresponding kinematic domain. To perform the fit, $\chi^2$ minimization technique~\cite{minuit2} as available within the  Minuit2 package of the ROOT ~\cite{root} software has been adopted. 
The fit results are depicted in Fig.~\ref{fig:fitting plot} and the corresponding values of $\sigma_{abs}$ along with the associated fit uncertainties, for different suppression scenarios, are listed in Table~\ref{tab:Fitting result}.~\footnote{The kinks in the model calculated suppression curves appear due to the discrete radius and diffuseness parameters of the Wood-Saxon nuclear density distribution used to evaluate the nuclear thickness function, $T_{A}(\vec{b})$ in the Glauber formalism employed to model the final state absorption (see Eq.~\ref{Eq:Full Glauber}). Explicit calculations show that the kinks in the curve disappear when instead of full Glauber approach, the so called exponential $<\rho L>$ parametrization ($S_{J/\psi} \sim$ $e^{-\rho_{0}\sigma_{abs}L}$) is adopted to model final state absorption. Also note that such kinks are less prominent when results are plotted in terms of $<L_{A}>$ which has much shorter range than the target mass number ($A$).} 
At 400 GeV, we have made a combined fit to both the NA50 and NA60 data sets. As evident from Table~\ref{tab:Expt. details}, the kinematic range of the  NA60 data lies entirely within the kinematic range of the NA50 measurements. Explicit calculation shows that separate fits to the NA50 and the  NA60 data sets produce $\sigma_{abs}^{J/\psi}$ values, which, within uncertainty, match the value obtained from the combined fit. This implies that the difference in the nuclear absorption, caused by the non-overlapping region in the kinematic ranges of the two measurements, is rather small and lies within the uncertainties of the data. In absence of any initial state energy loss effect, the extracted values of $\sigma_{abs}^{J/\psi}$ are larger than the corresponding $\sigma_{eff}^{J/\psi}$ values obtained from a pure Glauber model analysis~\cite{Chatterjee:2022ssu}. This feature can be attributed to the anti-shadowing effects in the primordial $c\bar{c}$ production caused by the employed nPDF scheme to model the parton densities inside the target, as was first reported in~\cite{Lourenco:2008sk}. Of course the absolute  value of the extracted $\sigma_{abs}$ for a given data set depends on the particular nPDF set. It is also interesting to note that inclusion of initial state energy loss leads to around $50 \%$ reduction in $\sigma_{abs}^{J/\psi}$, at all the investigated energies. As discussed before, energy loss of partons reduces the effective center-of-mass energy available for hard collisions and thus reduces the effective number of the  produced $c\bar{c}$ pairs. The reduction is strong enough to overcome the anti-shadowing effects due to enhanced parton densities inside the target nuclei, in the SPS energy domain. However, the absolute values of $\sigma_{abs}^{J/\psi}$, as extracted for the two models of energy loss, within uncertainty, agree with each other. This indicates that the target mass dependence of inclusive $J/\psi$ production data at the SPS facility is insensitive to distinguish the various scenarios of path length dependence of projectile parton energy loss. The small difference between the central values of the extracted $\sigma_{abs}^{J/\psi}$ parameter with relatively smaller values for BDMPS scheme can be understood from the nature of the shadowing function, where BH model has been seen to generate little lesser initial state suppression. The extracted $\sigma_{abs}^{J/\psi}$ values both in the presence and absence of the energy loss, exhibit a non-negligible dependence on the beam energy  with larger absorption at lower beam energies, in line with our present understanding.

\begin{figure*}
    \includegraphics[width=0.49\linewidth]{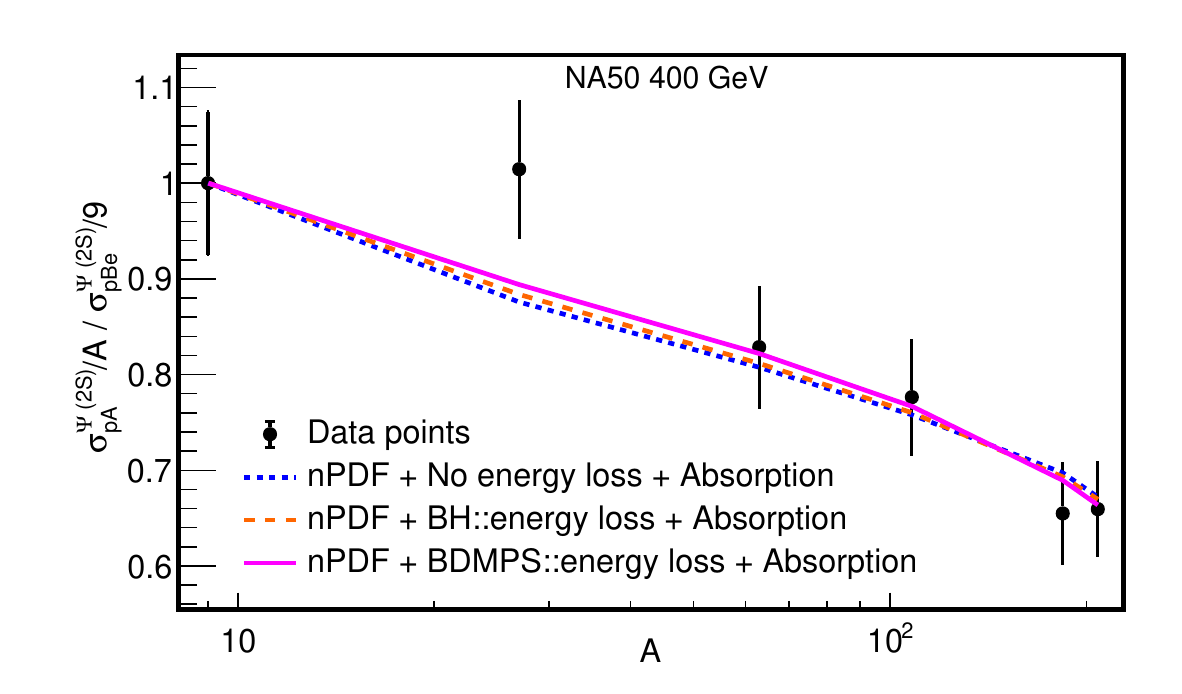}
    \includegraphics[width=0.49\linewidth]{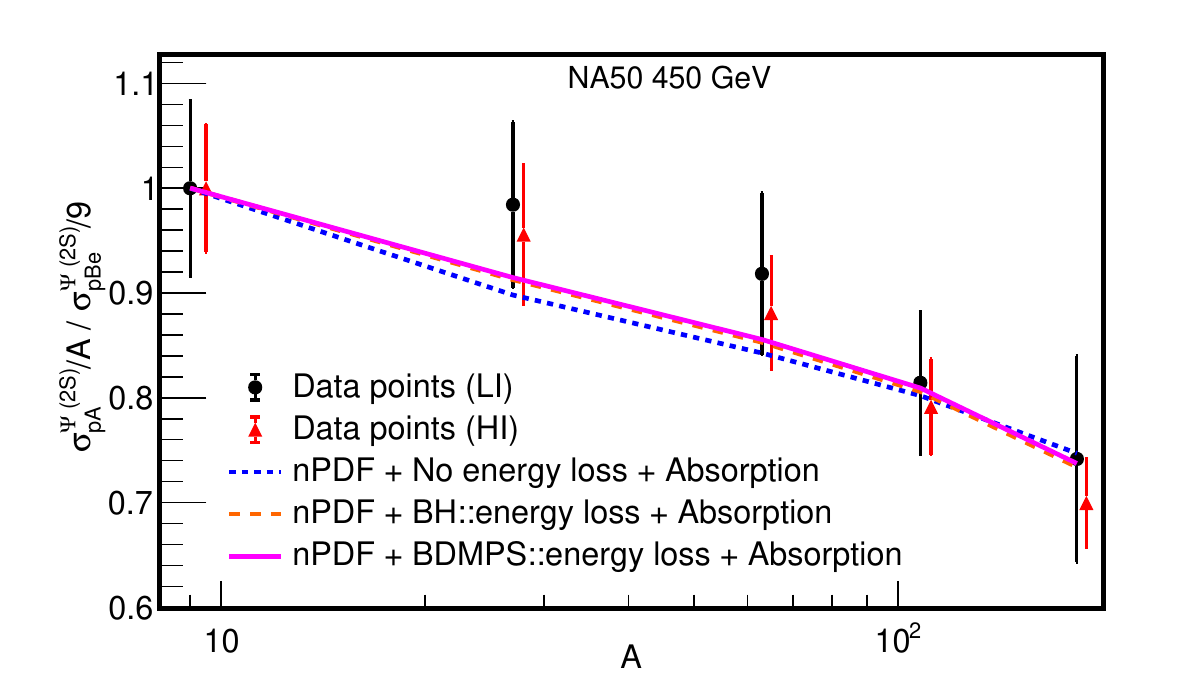}
    \caption{Ratio of the per nucleon $\psi (2S)$ production cross sections in p+A to p+Be collisions as a function of target mass (A) as measured at the SPS facility by the NA50 experiment at 400 GeV (left) and 450 GeV (right). At 450 GeV high intensity (HI) and low intensity (LI) data sets are separately shown. Three different lines correspond to the fitting of our model calculations to the available data for three different theoretical scenarios: nuclear modification of parton densities (nPDF) plus the final state absorption (dotted line), nuclear modification of parton densities (nPDF) along with incident parton energy loss following BH (dashed line) and BDMPS (solid line) schemes in presence of final state absorption. For each case the absorption cross section $(\sigma_{abs}^{\psi(2S)})$ has been extracted separately and listed in Table 3.}
    \label{fig: Fitting plot PsiP}
\end{figure*}

\begin{table}[htpb]
    \centering
    \begin{tabular}{|c|c|c|c|}
        \cline{1-4}
        Experiment & Energy loss & $\sigma^{\psi(2S)}_{abs}(mb)$ & $\chi^2/ndf$ \\ 
        & model &  & \\ \cline{1-4}
        \multirow{3}{*}{NA50 400 Gev}  & No   &  8.89$\pm$1.1  & 0.97   \\ 
        \cline{2-4}
          & BH   & 6.05$\pm$1   & 1.08   \\ 
          \cline{2-4}
           & BDMPS  & 5.80$\pm$1  & 0.88  \\ \cline{1-4}
           
        \multirow{3}{*}{\shortstack{NA50 450 GeV \\ (LI $\&$ HI)}} & No  & 7.06$\pm$0.88  & 0.60  \\ \cline{2-4}

          & BH  & 4.51$\pm$0.80   & 0.50  \\ \cline{2-4}
          & BDMPS  & 4.40$\pm$0.78  & 0.37  \\ \cline{1-4}
       
    \end{tabular}
     \begin{minipage}{0.91\columnwidth}
        \justifying 
    \caption{Extracted final state absorption cross section of $\psi(2S)$ mesons ($\sigma^{\psi(2S)}_{abs}$) obtained from the fit to the ratio of the per nucleon $\psi(2S)$  production cross sections measured by NA50 experiment in 400 GeV and 450 GeV fixed target $p+A$ collisions at the SPS, under various scenarios of initial state parton energy loss. A combined fit is performed for the HI and LI data sets at 450 GeV. } 
    \label{tab:Fitting result psi2s}
    \end{minipage}
\end{table}

We extend our investigations further by analyzing the $\psi(2S)$ production in 400 GeV and 450 GeV p+A collisions at the SPS facility as measured by the NA50 experiment. The fit results are displayed in Fig.~\ref{fig: Fitting plot PsiP} and the corresponding best fit parameters for $\sigma_{abs}^{\psi(2S)}$ are listed in Table~\ref{tab:Fitting result psi2s}. The final state absorption cross section for $\psi(2S)$ mesons is larger than that for $J/\psi$, which can be attributed to the larger size and weak binding of the $c\bar{c}$ pairs forming $\psi(2S)$ resonance states. For both the beam energies, $\sigma_{abs}^{\psi(2S)}$ is largest in absence of initial state parton energy loss, as expected. A pure Glauber model analysis without explicitly incorporating any initial state effect yields a value of $\sigma_{abs}^{\psi(2S)}$ around 8 mb. For two parametrizations of energy loss, alike $J/\psi$ results, the corresponding values of $\sigma_{abs}^{\psi(2S)}$ appear to be identical within errors. The stronger dissociation for $\psi(2S)$ states as compared to $J/\psi$ as observed in the data, certainly supports the nuclear absorption as the suitable scenario for the final state interaction. 

\begin{figure*}
\centering
    \includegraphics[width=0.49\linewidth]{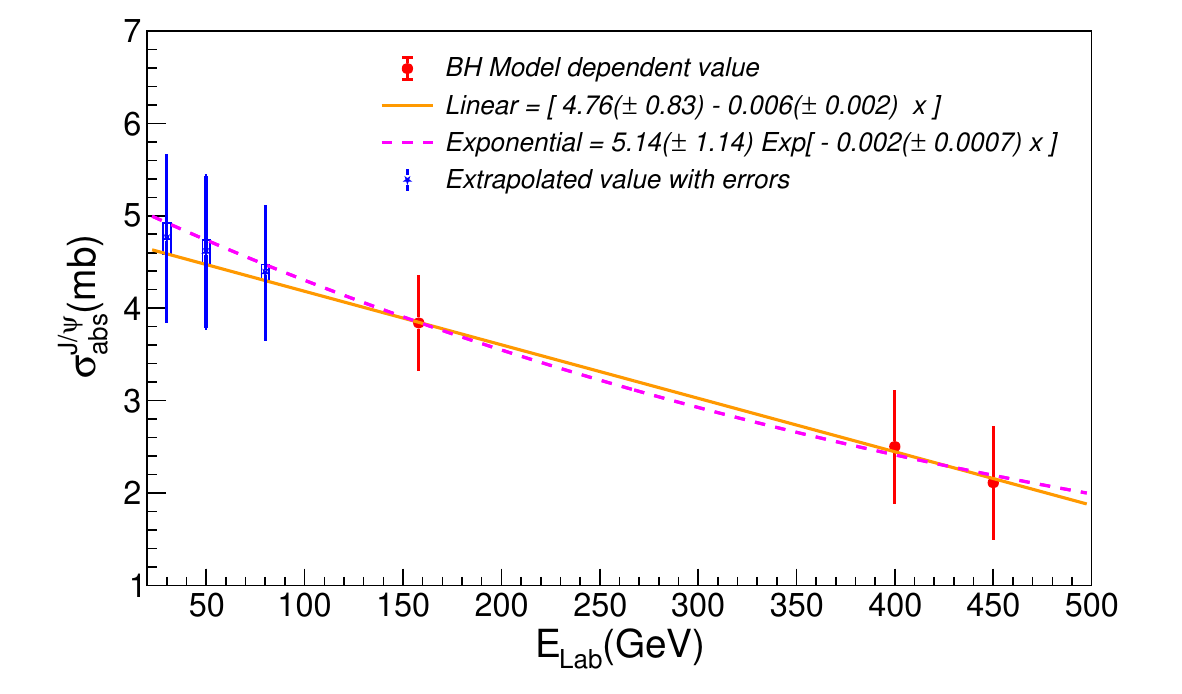}
    \includegraphics[width=0.49\linewidth]{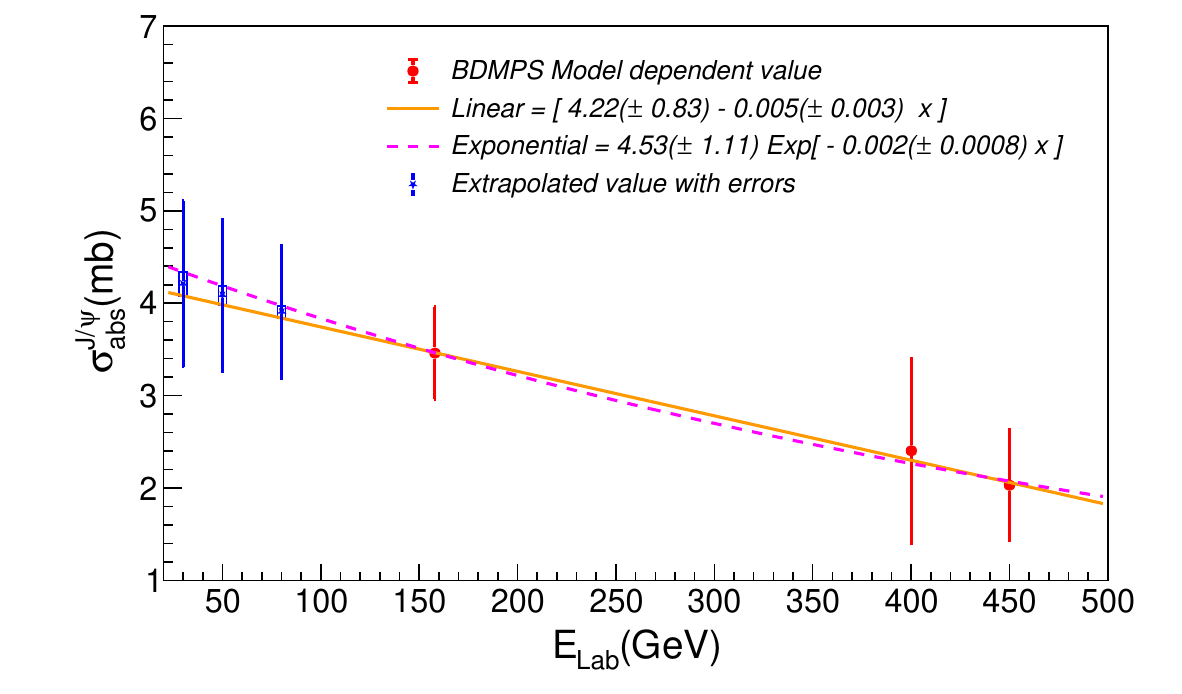}
    \includegraphics[width=0.49\linewidth]{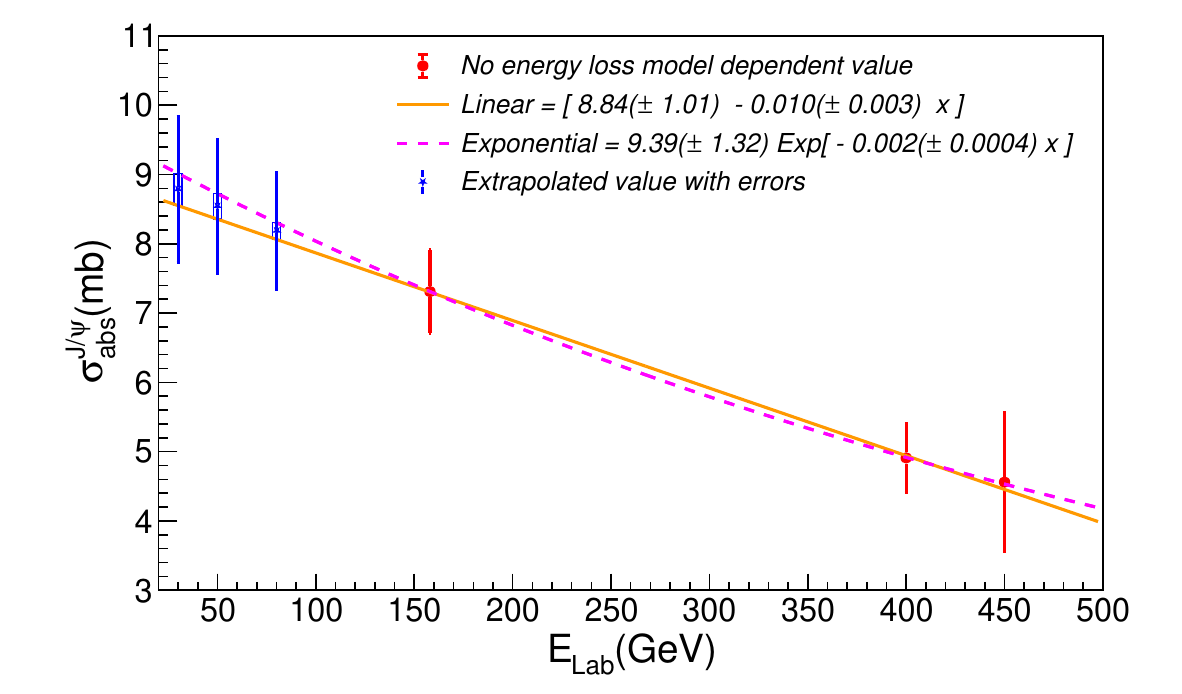}
    \caption{Parametrization of the beam energy dependence of the extracted values of $\sigma_{abs}^{J/\psi}$ fitted with two different functional forms. The solid (orange) line denotes a Linear function whereas the dashed (magenta) line is for Exponential function. $\sigma_{abs}^{J/\psi}$ values as extracted with BH scheme (left panel), BDMPS scheme (right panel) and no energy loss scenario (bottom panel) are shown separately.}
    \label{fig:extrapolation}
\end{figure*}

Our final goal is to predict the level of $J/\psi$ suppression expected in low energy p+A collisions in the upcoming experiments at SPS, J-PARC and FAIR.
\begin{table}[htpb]
    \centering
    \begin{tabular}{|c|c|c|c|c|}
    \cline{1-5}
      Energy loss & \multicolumn{4}{c|}{Paremeters of the fit functions }   \\
      \cline{2-5}
      
        model         &  \multicolumn{2}{c|}{Linear}   &   \multicolumn{2}{c|}{Exponential}   \\
      \cline{2-5}
          & $p_{0} (mb)$ & $p_{1} (mb.{GeV^{-1}})$ & $p_{0} (mb)$ & $p_{1} (GeV^{-1})$ \\
          \cline{1-5}
          \multirow{2}{*}{No} & 8.84 & - 0.01 & 9.39 & 0.002\\
          &  $\pm$ 1.01 & $\pm$ 0.003 & $\pm$ 1.32 & $\pm$ 0.0004\\
          \cline{1-5}   
        \multirow{2}{*}{BH } & 4.76 & - 0.006 & 5.14 & 0.002 \\
        &$\pm$ 0.83&$\pm$ 0.002&$\pm$ 1.14 &$\pm$ 0.0007 \\
       \cline{1-5}
       
        \multirow{2}{*}{BDMPS  } &4.22 &- 0.005&4.53&0.002 \\
       &$\pm$ 0.83 &$\pm$0.003&$\pm$ 1.11 &$\pm$ 0.0008\\
       \cline{1-5}
    \end{tabular}
    \begin{minipage}{0.98\columnwidth}
        \justifying
    \caption{Best fit parameters of the fitting functions (Linear and Exponential), used to parameterize the beam energy dependence of the $J/\psi$ final state absorption cross sections ($\sigma_{abs}^{J/\psi}$). See text, for details.}
    \label{tab:Parameters value}
    \end{minipage}
\end{table}

However, before doing so, it would be worth to discuss some limitations and uncertainties of our current study. We have used the leading order calculation for $c\bar{c}$ production within the framework of the color evaporation model (CEM). Next-to-leading order (NLO) processes, such as, one-loop gluon exchange, gluon emission, gluon splitting, and quark-gluon scattering are also important for charmonium production. As we are analyzing the ratio of production cross sections, the effect of NLO corrections to charm production would be small, if we assume the higher order contributions can be accounted by the multiplicative $K$-factor. More rigorous models, such as the Improved Color Evaporation Model (ICEM~\cite{Ma:2016exq}), Non-Relativistic Quantum Chromodynamics (NRQCD~\cite{Bodwin:1994jh}), and Modified NRQCD are also available in literature to describe charmonium production as well as their kinematic distribution in nuclear collisions. Since we are comparing inclusive charmonia production from light to heavy target nuclei, the choice of model for \(c\bar{c}\) pair production is likely to have a minimal impact, at least on the qualitative features of our obtained results. Fluctuations in the incoming parton energy loss has also been ignored in the present study. Another limitation arises from the yet undecided path length dependence of initial state parton energy loss. Future data on DY production in p+A collisions to be collected in the upcoming experimental facilities at the CERN SPS, JPARC in Japan, the CBM at FAIR SIS100, will be useful to clarify the issue. It is also worth mentioning that for all our calculations, we used the central set of parton densities based on the most probable values of the EPPS21 NLO nuclear parton distribution function (nPDF) parameters. Like other nPDFs, EPPS21 contains substantial uncertainty bands that reflect how effectively the parameters are fixed from the underlying data. Unlike its predecessors (EPS09 or EPPS16), EPPS21 explores uncertainties in the nuclear PDFs originating from the baseline free proton PDFs by propagating associated errors within the Hessian framework. The EPPS21 nPDF scheme includes a total of 106 error sets, where sets 1-48 correspond to uncertainties from nuclear modifications and sets 49-106 account for uncertainties in the baseline proton PDFs (CT18ANLO). We checked our extracted $\sigma_{abs}^{J/\psi}$ values against the uncertainty set, to find that they change by about 5\%, at all the investigated energies.
On the other hand, for a quantitative estimation of the uncertainty present in our results originating from the combination of the NLO PDFs with LO processes, we have carried out the following check. Instead of EPPS21 nPDF sets and CT18NLO free proton PDF sets, we use EPS09 nPDF package which is available for both LO and NLO sets. DY and $c\bar{c}$ production cross sections in p+A collisions are thus calculated independently with (i) EPS09 LO nPDF set combined with MSTW 2008 LO free proton PDF set as well as with (ii) EPS09 NLO nPDF set combined with MSTW2008 NLO free proton densities. The resulting difference in the initial state quark energy loss parameter $\alpha$ ($\beta$) is  $4 \%$ ($5\%$) for the BH (BDMPS) scheme of energy loss, as obtained by analyzing DY data from Fermilab E906 experiment in 120 GeV p+A collisions. With the updated values of $\alpha$ ($\beta$) when we analyze the NA60 data on $J/\psi$ production cross sections in 158 GeV p+A collisions, the corresponding final state absorption cross section with BH (BDMPS) scheme comes out to be $\sigma_{abs}^{J/\psi} = 5.12 \pm 0.6$ mb ($ 4.73 \pm 0.8$ mb) with LO parton densities and $\sigma_{abs}^{J/\psi} = 4.21 \pm 0.6$ mb ($ 3.78 \pm 0.8$ mb) with NLO parton densities. Thus around $ 17 - 18 \%$ difference in the $\sigma_{abs}^{J/\psi}$ parameter is expected to originate from the higher order corrections built in the opted PDF sets. Such variations can be attributed to the difference in the associated gluon densities in the employed PDF sets. On the other hand, adoption of $<\rho L>$ parametrization, instead of the full Glauber model, leads to a change in the corresponding values of $\sigma_{abs}^{J/\psi}$ below $10\%$. One may also note that at 450 GeV we have made a combined fit to HI and LI data sets. A separate fit to HI and LI data sets would generate $J/\psi$ absorption cross sections which are $ 3- 5 \% $ larger than the common value. Similarly, instead of a combined fit, if we separately fit the NA50 and NA60 data at 400 GeV, the resulting $\sigma_{abs}^{J/\psi}$ values deviate by $5 - 6 \%$  from the common value. On the other hand, in lieu of the ratio, if we fit the mass dependence of the absolute production cross sections, then the extracted $\sigma_{abs}^{J/\psi}$ values vary less than $5\%$ than those listed in Table~\ref{tab:Fitting result}. Considering all the above sources of systematics to be uncorrelated, an overall  uncertainty less than $20 \%$  is expected on our estimated values of $\sigma_{abs}^{J/\psi}$.

Finally, before moving forward one may note that in our calculations we have assumed that $\sigma_{abs}^{J/\psi}$ remains constant throughout the transit of the target. However there are some evidences in literature, where $c\bar{c}$-nucleon disintegration cross section is anticipated to depend strongly on the size of the evolving $c\bar{c}$ pair as it expands to a fully formed physical resonance and thus on $\tau$, the average time $c\bar{c}$ spends inside the nucleus~\cite{Farrar:1990ei,Blaizot:1989de,Gavin:1990gm,Arleo:1999af,McGlinchey:2012bp}. Following this prescription of time dependent nuclear absorption, as modeled in Ref.~\cite{Arleo:1999af}, the authors of Ref.~\cite{McGlinchey:2012bp} have fitted the centrality and rapidity dependent $J/\psi$ data from PHENIX experiment in $\sqrt{s_{NN}} =200 $ GeV d+Au collisions, using Glauber model with an effective absorption cross section, $\sigma_{abs}$. The extracted values of $\sigma_{abs}$, from the PHENIX data at $\sqrt{s_{NN}} =200$ GeV and from lower energy fixed target collision data with $17.3 < \sqrt{s_{NN}} < 41.6$ GeV, are found to display a common scaling behavior of $\sigma_{abs}$ with $\tau$, for $\tau > 0.05$ fm/$c$. The dependencies of $\sigma_{abs}$ on impact parameter of the collision and on the target mass are found to be small. The best fit values of $\sigma_{abs}$ versus $\tau$ are found to provide an excellent description of the beam energy dependence of nuclear absorption at mid-rapidity in the collision energy range, $\sqrt{s_{NN}} = 20 - 200$ GeV. Incorporation of time dependence in final state nuclear dissociation however requires additional parameters which need to be fixed from the data. Explicit calculations show that, in the kinematic domain probed by our analyzed data sets and for an intrinsic resonance formation time of $\tau_{0} = 0.25$ fm/$c$, $J/\psi$ mesons remain around $65 - 85 \%$ of the total time spent inside a heavy target ($A=200$) as fully formed resonances. Our estimated values of $\sigma_{abs}^{J/\psi}$ can thus be acceptable and considered equivalent to averages over the entire time taken by the nascent $c\bar{c}$ pairs, in their pre-resonant and resonant state, to exit the target.

\begin{table}[htpb]
    \centering
    \begin{tabular}{|c|c|c|}
        \cline{1-3}
        Lab Energy & Energy loss model & $\sigma^{J/\psi}_{abs}(mb)$  \\ \cline{1-3}
        \multirow{3}{*}{80 GeV}  & No   &  8.16$\pm$0.61    \\ 
        \cline{2-3}
          & BH   & 4.36$\pm$0.5      \\ 
          \cline{2-3}
           & BDMPS  & 3.90$\pm$0.5   \\ \cline{1-3}
           
        \multirow{3}{*}{50 GeV} & No  & 8.51$\pm$0.7  \\ 
        \cline{2-3}

          & BH  & 4.56$\pm$0.6     \\ 
          \cline{2-3}
          & BDMPS  & 4.07$\pm$0.6   \\ 
          \cline{1-3}

          \multirow{3}{*}{30 GeV} & No  & 8.74$\pm$0.8  \\ 
        \cline{2-3}

          & BH  & 4.72$\pm$0.6     \\ 
          \cline{2-3}
          & BDMPS  & 4.19$\pm$0.6   \\ 
          \cline{1-3}
       
    \end{tabular}
     \begin{minipage}{0.86\columnwidth}
        \justifying 
    \caption{The predicted values of $J/\psi$ absorption cross section $\sigma^{J/\psi}_{abs}$ for different projectile parton energy loss parameterizations at the beam energies of 80 GeV, 50 GeV and 30 GeV as will be available from the future fixed target $p+A$ collision experiments. } 
    \label{tab:Extrapolated result}
    \end{minipage}
\end{table}

\begin{figure*}
\centering
    \includegraphics[width=0.49\linewidth]{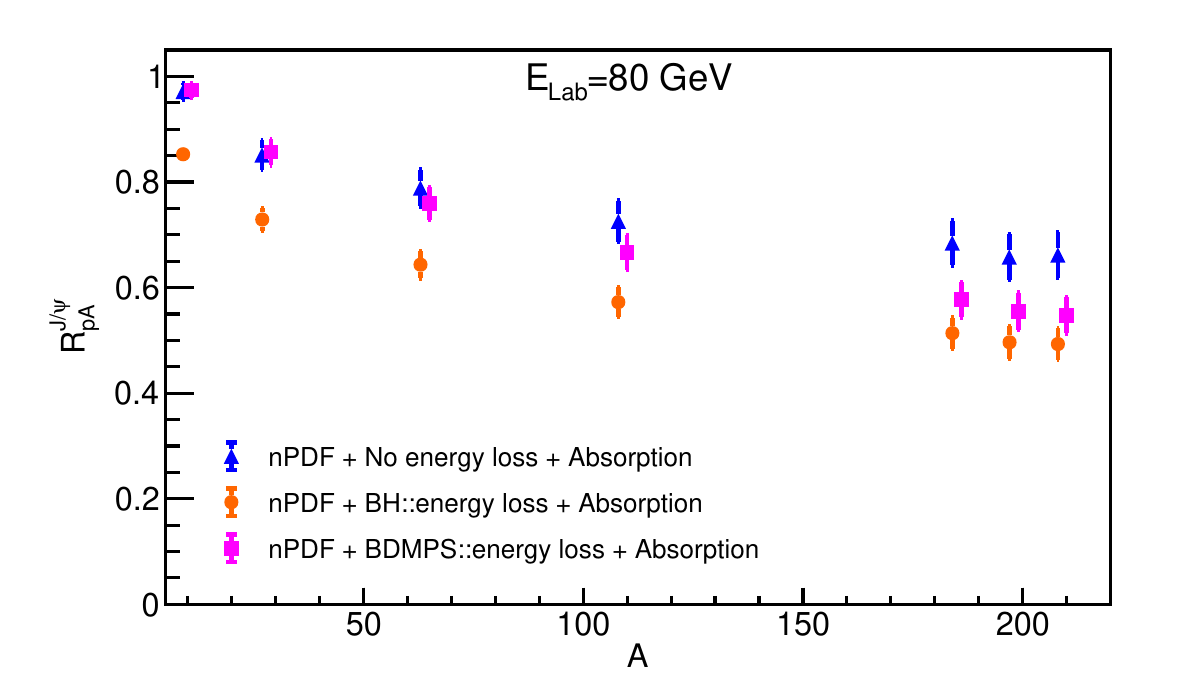}
    \includegraphics[width=0.49\linewidth]{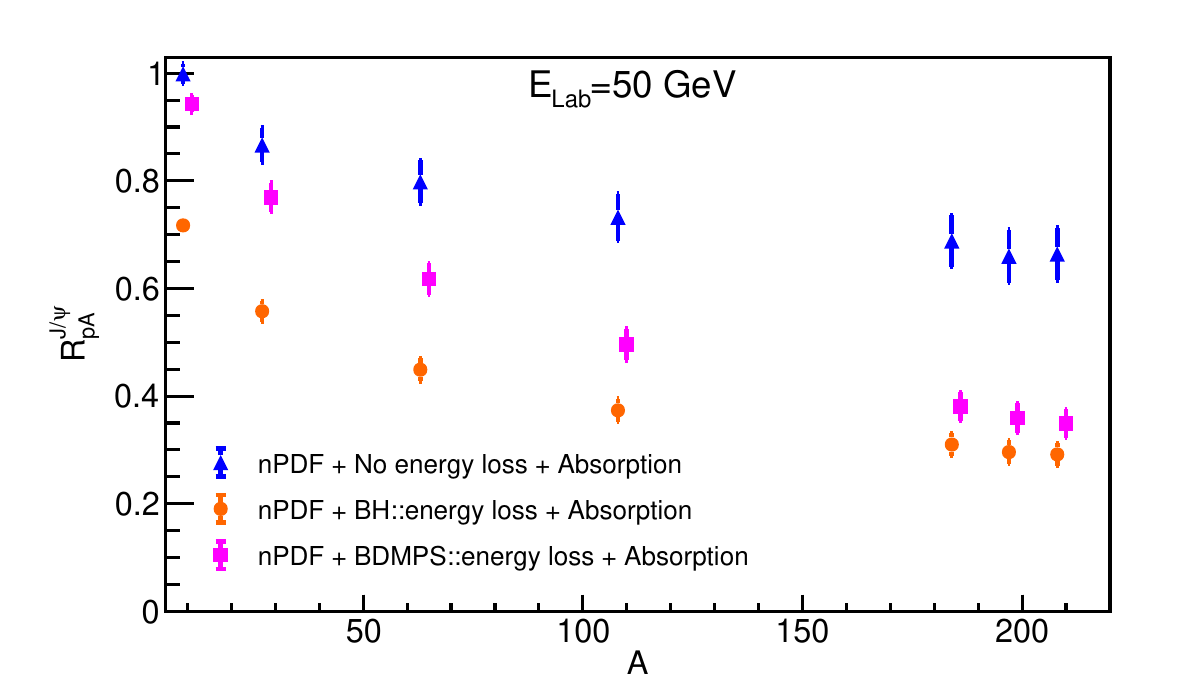}
    \includegraphics[width=0.49\linewidth]{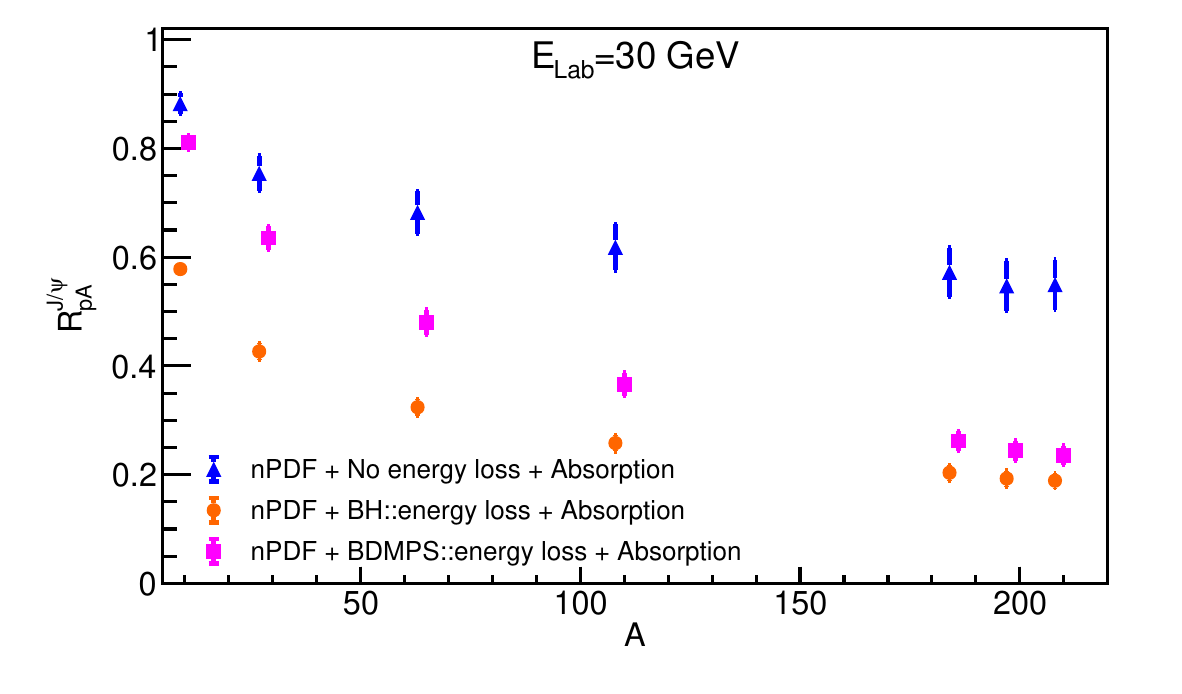}

     \caption{The predicted level of $J/\psi$ suppression in the future fixed target p+A collision experiments. Results are expressed in terms of the ratio of the per nucleon J/$\psi$ production cross sections in p+A collisions to that in p+p collisions, as a function of target mass ($A$), for the incident proton beam energies of 80 GeV, 50 GeV and 30 GeV. The model calculations are presented for three different scenarios of CNM dissociation: nuclear modification of parton distribution inside target (nPDF) at the initial state associated with final state absorption, nPDF coupled to beam parton energy loss following BH and BDMPS formalisms in the initial state associated with final state absorption. The error bars associated to the simulated data points originate from the corresponding fit uncertainty associated with the estimation of the $\sigma^{J/\psi}_{abs}$. Variation in the mean value of initial state energy loss parameter will induce additional $5 - 9 \%$ systematic uncertainty in the estimated values of the production cross section, common to all nuclear targets.}
    \label{fig:Prediction_future_Experiments}
\end{figure*}

We conclude this section by discussing the expected level of CNM suppression on charmonium production in upcoming fixed-target p+A collisions at various accelerator facilities. To date, the lowest proton beam energy used is 158 GeV at CERN SPS, where inclusive $J/\psi$ production was measured by the NA60 experiment for a variety of nuclear targets.  High intensity proton beam of energy 50 GeV will be available from the Japan Proton Accelerator Research Complex (J-PARC), while $J/\psi$ production in p+A collisions will be investigated by the NA60+ muon spectrometer at SPS within the energy range of 40-160 GeV. At FAIR SIS00 facility, the CBM experiment also plans to measure charm production in 30 GeV p+A collisions. Such low energy collisions will typically probe a phase space region $(0.2<x_{2}<0.5)$ where parton densities inside the target would exhibit from weak anti-shadowing to eventual shadowing effects. The beam parton energy loss, on the other hand, increases with decreasing beam energy, leading to stronger suppression of initial state $c\bar{c}$ production even in the absence of any final state dissociation. Measurement of hard probes in low energy collisions thus provides a unique opportunity for a comprehensive investigation of the relative interplay of different initial state CNM effects. However,  the overall suppression pattern of $J/\psi$ production in p+A collision will also depend on the magnitude of final state absorption of the nascent resonant states inside the nuclear matter. As evident from Fig.~\ref{fig:Formation Length}, the variation of the formation length, in the kinematic domain, that will be probed by these future experiments, is entirely consistent with the final state nuclear absorption scenario. Hence, to predict the degree of nuclear absorption in these low energy collisions, we first parametrize the beam energy dependence of the extracted absorption cross sections and extrapolate to the lower collision energies. In Fig.~\ref{fig:extrapolation}, we fit the beam energy dependence of \(\sigma^{J/\psi}_{\text{abs}}\).
Two different functional forms, namely, a Linear ($\sigma_{abs}^{J/\psi}(E_{Lab}) =p_{0} + p_{1}E_{Lab}$) and an Exponential ($\sigma_{abs}^{J/\psi}(E_{Lab})=p_{0}e^{{- p_{1}E_{Lab}}}$) are employed\footnote{Such parametrizations are valid above the kinematic threshold for $J/\psi$ production.} to perform the fit and the corresponding fit parameters are listed in Table \ref{tab:Parameters value}.
The expected \(\sigma^{J/\psi}_{\text{abs}}\) values are calculated separately, at three representative beam energies: 80 GeV, 50 GeV and 30 GeV, typical to the three future experimental programs. These values so obtained for the two parametrizations are averaged to reduce the systematic uncertainty in the estimation of \(\sigma^{J/\psi}_{\text{abs}}\) and are listed in Table~\ref{tab:Extrapolated result}. As evident, the $\sigma_{abs}^{J/\psi}$ in absence of any kind of initial state energy loss effect, comes out to be almost twice as large as compared to its presence. Two models of initial state energy loss give comparable values of $\sigma_{abs}^{J/\psi}$ within errors. Though small but increasing difference between the central values of $\sigma_{abs}^{J/\psi}$ with decreasing beam energy is due to the growing difference in the initial state suppression patterns between the two models (see Fig.~\ref{fig:S_vs_L}). The resultant suppression pattern as a function of target mass number is illustrated in Fig.~\ref{fig:Prediction_future_Experiments}, in terms of the observable nuclear modification factor ($R_{pA}^{J/\psi}$), defined as the ratio of per nucleon $J/\psi$ production cross sections in p+A and p+p collisions. The required cross section for p+p collisions can in principle be measured using a liquid hydrogen target and will also be useful to estimate nuclear modification factor ($R_{AA}$) in ion-ion collisions. Presence of initial state energy loss produces a stronger overall suppression pattern, even with an almost two times weaker final state absorption. Difference between the suppression patterns in presence and  absence of the energy loss effects grows larger with decreasing beam energy. However the two parameterizations of initial state energy loss will not be distinguishable by measurement of $J/\psi$ production alone.

.

\section{Summary}
Identifying the suppression of $J/\psi$ mesons as a unique signature of QGP formation in relativistic nuclear collisions requires a precise and quantitative understanding of various other dissociation mechanisms unrelated to the deconfinement transition. CNM effects have long been recognized as a significant source of $J/\psi$ suppression in both p+A and A+A collisions. Since data from heavy-ion collisions are influenced by a combination of cold and hot medium effects, isolating the genuine QGP-induced suppression solely from heavy-ion data is challenging. Hence, p+A collisions provide a crucial baseline for quantifying the impact of CNM effects on $J/\psi$ production. Multiple effects arising from different physical origins and acting at various stages of resonance formation are collectively referred to as CNM effects. These effects lead to an overall suppression of charmonium production in p+A collisions. CNM effects can be broadly categorized into two classes: initial-state effects, which occur before the hard scattering and influence the production of the $c\bar{c}$ pairs, and final-state effects, which affect the hadronization of the evolving $c\bar{c}$ pairs into physical resonance states. The initial-state effects include the modification of parton distribution functions within the target nucleus, together with the energy loss experienced by the incoming beam partons as they traverse the nuclear medium. The final-state effects, occurring after the production of the $c\bar{c}$ pair, involve either the nuclear absorption of the pre-resonant or resonant charm-quark pair, or the energy loss of the colored $c\bar{c}$ system through soft gluon radiation while propagating through the target. \\

In this study, we focus on the experiments that demonstrate nuclear absorption effects in the final state, rather than final-state energy loss. The nuclear modifications to parton distribution functions are implemented using the EPPS21 NLO nPDF set in conjunction with the CT18A NLO free-proton PDFs, both obtained from the state-of-the-art LHAPDF framework. The beam parton energy loss inside the target is quantified by fitting the nuclear DY data from the Fermilab E866 and E906 experiments, which studied di-muons produced in 800 GeV and 120 GeV proton beams incident on various nuclear targets~\cite{Giri:2025bfq}.
The NA50 and NA60 experiments at the CERN SPS facility have investigated the CNM effects on $J/\psi$ production using proton beams with energies between 450 GeV and 158 GeV across a broad range of target nuclei. In this work, we perform a leading-order phenomenological analysis of the $J/\psi$ production cross sections and their ratios as measured by these experimental collaborations. The analysis systematically accounts for all established initial state effects on charmonium production cross sections to constrain the relevant cold nuclear matter effects. In this study, we excluded the experimental results from E866, E906 and HERA-B on charmonium production due to their formation outside the nuclear medium. We present revised results obtained from a re-evaluation of charmonium production data from the NA50 and NA60 experiments, which reveal multiple nuclear effects contributing to the observed $J/\psi$ suppression. Our analysis highlights the critical role of energy loss in this suppression mechanism. These findings motivate a quantitative reexamination of charmonium dissociation in the deconfined quark–gluon plasma (QGP) phase, guided by heavy-ion collision data.\\

Our systematic analysis reconfirms the previously observed pronounced beam-energy dependence of the nuclear absorption cross section for $J/\psi$ mesons in fixed-target p+A reactions. Upon incorporating the initial-state parton energy loss, the extracted $\sigma^{J/\psi}_{\text{abs}}$ value is found to decrease significantly. This reduction is expected to be even more pronounced for higher charmonium states such as $\psi(2S)$ and $\chi_c$. However, limited statistics for $\psi(2S)$ and no results for $\chi_{c}$ at the existing fixed-target energies constrain detailed investigations. Nevertheless, the increased luminosity anticipated at upcoming experimental facilities offers promising prospects for precise measurements of these states. The extracted $\sigma_{abs}^{J/\psi}$ is interpreted purely as a manifestation of final-state nuclear absorption, disentangled from the other two key initial-state cold matter effects — modified momentum distribution of target partons as well as the energy loss of projectile partons. This analysis provides a determination of the $J/\psi$ final-state absorption cross section independent of initial-state influences. However, some residual contributions from the nuclear effects on the higher charmonium states $\chi_c$ and $\psi(2S)$ may still propagate into the extracted $\sigma_{abs}^{J/\psi}$, as approximately 33\% of the observed $J/\psi$ yield originates from their decays. These feed-down states experience stronger nuclear absorption than directly produced $J/\psi$. 
Finally, we offer predictions for the expected magnitude of nuclear suppression of $J/\psi$ mesons at fixed-target beam energies of 80, 50, and 30 GeV — relevant for upcoming experimental programs at the CERN SPS, J-PARC, and CBM-FAIR, by extrapolating the results of our existing analyses. The upcoming experimental facilities will explore a kinematic regime of charmonium production where nuclear shadowing effects are expected to be small. However presence of initial state beam parton energy loss coupled with growing magnitude of final state absorption will lead to enhanced CNM suppression of $J/\psi$ production in low energy p+A collisions as compared to the available results. For the heavy-ion collisions the influence of CNM effects are likely to become stronger with decreasing beam momentum as the two colliding nuclei will take longer time to pass through each other. Dedicated measurements of $J/\psi$ production in p+A collisions in the same energy and kinematic domain of the ion-ion collisions will thus be essential for a quantitative analysis of the heavy-ion data.

\begin{acknowledgements}
The authors are thankful to F. Arleo, E. G. Ferreiro and Sushant K. Singh for many useful discussions. SKD and PPB acknowledge the support from DAE-BRNS, India, Project No. 57/14/02/2021-BRNS.
\end{acknowledgements}

\end{document}